\documentclass[preprintnumbers,nofootinbib,unsortedaddress,superscriptaddress,notitlepage]{revtex4-1}
\usepackage{amsfonts}
\usepackage{amsmath}
\usepackage{hyperref}
\usepackage{amssymb}
\usepackage[english]{babel}
\usepackage{graphicx}
\usepackage{epsfig}
\usepackage{bm}
\usepackage{longtable}
\usepackage{verbatim}
\usepackage{longtable}
\usepackage{cleveref}
\usepackage{float}
\usepackage{subfigure}
\usepackage[utf8]{inputenc}
\usepackage{xcolor}

\input{tcilatex}
\begin{document}

\title{Constraints on vector resonances from a strong Higgs sector.}

\author{A. E. C\'arcamo Hern\'andez}
\email{antonio.carcamo@usm.cl}
\author{Basti\'an D\'iaz S\'aez}
\email{bastian.diaz@alumnos.usm.cl}
\author{Claudio O. Dib}
\email{claudio.dib@usm.cl}
\author{Alfonso Zerwekh}
\email{alfonso.zerwekh@usm.cl}
\affiliation{{\small Department of Physics and Centro Cient\'{\i}fico-Tecnol\'ogico de
Valpara\'{\i}so}\\
Universidad T\'ecnica Federico Santa Mar\'{\i}a, Valpara\'{\i}so, Chile}

\begin{abstract}

We consider a scenario of a composite Higgs arising from a strong sector.
We assume that the lowest lying composite states are the Higgs scalar doublet and a
massive vector triplet, whose dynamics below the compositeness scale are
described in terms of an effective Lagrangian. Electroweak symmetry breaking takes place through a vacuum expectation value just as in the Standard Model, but with the vector resonances strongly coupled to the Higgs field.
We determine the constraints
on this scenario imposed by 
(i) the Higgs diphoton decay rate, 
(ii) the electroweak precision tests and
(iii) searches of heavy resonances at the LHC in the final states $l^+l^-$ and $l\nu_l$ ($l=e,\mu$), $\tau^+\tau^-$, $jj$, $t\bar{t}$, $WZ$, $WW$, $WH$ and $ZH$. We find that the heavy vector resonances should have masses that are constrained to be in the range $2.1$ - $3$ TeV. 
On the other hand, the mixing of the heavy vectors with the Standard Model gauge bosons is 
constrained to be in the range $\tan\vartheta\sim 0.1 - 0.3$, which is consistent with the assumption that the Higgs couples weakly  to the Standard sector, even though it couples strongly to the heavy vector resonances.

\end{abstract}

\maketitle

\section{Introduction.}

\label{intro} The recent discovery of the Higgs boson at the LHC \cite%
{Aad:2012tfa,Chatrchyan:2012xdj} provides the opportunity to directly
explore the mechanism of electroweak symmetry breaking (EWSB). While this
remarkable achievement implies severe constrains on many proposed extensions
of the Standard Model (SM), an additional sector beyond our current knowledge is
still needed in order to explain the dynamical origin of the electroweak
scale and its stability \cite{Agashe:2014kda}. A specific question in this
context is whether this new sector is weakly or strongly interacting \cite%
{Contino:2009ez}. In the latter case, the Higgs boson is viewed as a
composite state which must be accompanied by a plethora of new heavy
composite particles \cite{Grojean:2009fd,Contino:2010rs,Panico:2015jxa}. In
general, it is expected that the lightest states produced by the strong
dynamics would correspond to spin-0 and spin-1 particles \cite%
{Grojean:2009fd,Contino:2010rs,Panico:2015jxa,Arbey:2015exa}. In these models the
lightness of the Higgs can be explained in two different ways. One way is to
consider the Higgs boson as a pseudo-Goldstone boson that appears after the
breakdown of a suitable global symmetry \cite%
{Agashe:2004rs,Gripaios:2009pe,Contino:2010rs,Panico:2015jxa,Barbieri:2007bh,Csaki:2008zd,Contino:2011np,Mrazek:2011iu,Pomarol:2012qf,Contino:2013kra,Pappadopulo:2013vca,Montull:2013mla,Cacciapaglia:2014uja,Carena:2014ria,vonGersdorff:2015fta,Belyaev:2015hgo,Cacciapaglia:2015eqa,Fichet:2016xvs,Fichet:2016xpw,Ma:2017vzm}%
. A second way is to consider the Higgs boson as the modulus of an effective 
$SU(2)$ doublet, where its lightness is due to particularities of the
dynamics of the underlying theory \cite%
{Bardeen:1989ds,Lane:2005vp,Giudice:2007fh,Zerwekh:2005wh,Bai:2008gm,Lane:2009ct,Zerwekh:2010uk,Hernandez:2010iu,Hernandez:2010qp,Hernandez:2011rw,Burdman:2011fw,Eichten:2012qb,CarcamoHernandez:2012xy,Bellazzini:2012tv,Contino:2013kra,Diaz:2013tfa,Castillo-Felisola:2013jua,Hernandez:2013zho,Carcamo-Hernandez:2013ypa,Hernandez:2015xka,Pappadopulo:2014qza,Lane:2014vca,Lane:2015fza,Lane:2016kvg,Gintner:2016bhn,Foadi:2007ue,Ryttov:2008xe,Sannino:2009za,Belyaev:2013ida,Hapola:2011sd,Foadi:2012bb,Chala:2017sjk}%
. For instance, there are evidences that quasi-conformal strong interacting
theories such as walking technicolor may provide a light composite scalar 
\cite%
{Foadi:2007ue,Ryttov:2008xe,Sannino:2009za,Belyaev:2013ida,Hapola:2011sd,Foadi:2012bb}%
. It has also been shown that, in the effective low energy theory, the
composite scalar may develop a potential that reproduces the standard Higgs
sector \cite{Bardeen:1989ds}. In this scheme, the electroweak symmetry
breaking is effectively described by a non zero vacuum expectation value of
the scalar arising from the potential, just as in the Standard Model.
However, additional composite particles, like vector resonances, may also be
expected to appear in the spectrum \cite{Dietrich:2005jn}.

 The main reason to consider strongly interacting mechanisms of EWSB 
as alternatives to the Standard Model mechanism
based on a fundamental scalar is the so called hierarchy problem that arises
from the Higgs sector of the SM \cite%
{Grojean:2009fd,Contino:2010rs,Panico:2015jxa}. This problem is indicative
that, in a natural scenario, new physics should appear at scales not much
higher than the EWSB scale, say around a few TeV, in order to stabilize the
Higgs mass at a value much lower than the Planck scale ($\sim 10^{19}$ GeV).
An underlying strongly interacting dynamics without fundamental scalars,
which becomes non-perturbative somewhere above the EW scale, is a possible
scenario that gives an answer to this problem.

 While a composite Higgs boson is theoretically attractive because the
underlying strong dynamics provides a comprehensive and natural explanation
for the origin of the Fermi scale \cite%
{Grojean:2009fd,Contino:2010rs,Panico:2015jxa}, the presence of additional
composite states such as the vector triplets previously mentioned may, in
principle, produce phenomenological problems. For instance one could expect
that, at one loop level, they may produce sensible corrections to
observables involving the Higgs boson. Consequently, an interesting quantity
which can eventually reveal the influence of additional states is $\Gamma
(h\rightarrow \gamma \gamma )$. In a previous work this decay channel was
studied in a simple model with vector resonances and found it is in general
agreement with current experimental measurements in the limit where the
Higgs boson is weakly coupled to the new resonances \cite%
{Castillo-Felisola:2013jua}. 
However, if the Higgs boson arises from a
strongly interacting sector together with other heavy resonances, one should
expect a strong coupling among them. 

In this work, we want to investigate whether this
strong coupling hypothesis is still compatible with the currently known
phenomenology and, in general, whether composite models are viable
alternatives to electroweak symmetry breaking, given the current
experimental success of the Standard Model \cite{Agashe:2014kda}. 
To be concrete, we describe the new sector by means
of an effective model with minimal particle content, without referring to
the details of the underlying strong dynamics. We use an effective chiral
Lagrangian to describe the theory below the cutoff scale of the underlying
strong interaction, assumed to be $\Lambda =4\pi v\sim 3$ TeV. This low
energy effective theory must contain the Standard Model spectrum and the
extra composite scalar and vector multiplets. 

The content of this paper goes as follows. In section \ref{Model} we
introduce our effective Lagrangian that describes the spectrum of the
theory. Section \ref{Higgsdiphotonrate} deals with the constraints arising from the
Higgs diphoton decay rate and dijet exclusion limits. The constraints on the model parameter space arising from the oblique $T$ and $S$ parameters are discussed in Section \ref{TandS}. 
In section \ref{Decaychannels} we describe the different decay channels of the heavy vector
resonances. In section \ref{LHCsearches} we present the constraints of our model arising from LHC searches of heavy vector resonances. Finally, in section \ref{conclusions} we state our conclusions.

\section{Lagrangian for a Higgs doublet and heavy vector triplet.}

\label{Model} 
We want to formulate the scenario of EWSB
triggered by a strongly coupled sector without referring to specific details
of the underlying theory. This underlying theory, as it becomes strong at
low energies, should generate the Higgs scalar multiplet as a composite
field below a scale $\Lambda\sim 4\pi v$, where $v$ will be analog of the
pion decay constant in QCD. We will assume that, in addition to the
composite scalar multiplet, there will remain a vector composite multiplet below the scale 
$\Lambda$. One should then expect that these composite fields would exhibit
a remnant strong coupling among themselves, which is the main hypothesis we
want to test.

We will assume the vector composites to form  a triplet
under $SU(2)_L$, while the scalars will form a Higgs doublet just as in the SM. To this end we construct the effective theory based on a hidden local $SU(2)$ symmetry, so that our gauge group appears as $SU(2)_{1}\times SU(2)_{2}\times
U(1)_{Y}$. To give large masses to the vectors, the $SU(2)_1\times SU(2)_{2}$ part will be broken down to the
diagonal subgroup, i.e the standard $SU(2)_L$. The would-be
Goldstone bosons of this breaking will be incorporated as a non-linear sigma
model field $\Sigma $. In turn, the gauge symmetry of the Standard Model
will be broken, as usual, when the electrically neutral component of the
scalar doublet $\Phi$ acquires a vacuum expectation value.

We denote the gauge fields of $SU(2) _{1}$, $SU(2)
_{2}$ and $U\left( 1\right) _{Y}$ as $A^{(1)}_{\mu }$, $A^{(2)}_{\mu }$ and $B_{\mu }$%
, respectively. After the breaking of $SU(2)_{1}\times SU(2)_{2}\rightarrow
SU(2)_{L}$, one combination of the vector fields $A^{(1)}_{\mu }$ and $A^{(2)}_{\mu }$
will become the heavy vectors and the other combination will remain as the $%
SU(2)_{L}$ gauge fields. In our notation, the heavy vectors will be mainly $A^{(2)}_{\mu }$ 
with a small admixture of $A^{(1)}_{\mu }$. The scalar doublet $\Phi $%
, i.e. the Higgs field for the SM, on the other hand, should be completely localized at the $SU\left(
2\right) _{2}$ site, in order to reflect a stronger coupling to the heavy
vectors. As such, $\Sigma$ is a doublet under both $SU(2)_{1}$ and $SU(2)_{2}$,
while $\Phi $ and $\psi _{iL}$ are doublets only under $SU(2)_{2}$ and $%
SU(2)_{1}$, respectively (see Table \ref{tabla1}). The effective Lagrangian is expressed as

\begin{eqnarray}
\tciLaplace &=&-\frac{1}{2}\left\langle F^{(1)}_{\mu \nu }F^{{(1)}\mu \nu
}\right\rangle -\frac{1}{2}\left\langle F^{(2)}_{\mu \nu }F^{(2)\mu \nu
}\right\rangle -\frac{1}{2}\left\langle B_{\mu \nu }B^{\mu \nu
}\right\rangle +\frac{{f_{\Sigma }}^{2}}{2}\left\langle \left( D_{\mu
}\Sigma \right) ^{\dag }D^{\mu }\Sigma \right\rangle +\left( D_{\mu }\Phi
\right) ^{\dag }D^{\mu }\Phi -\mu ^{2}\Phi ^{\dag }\Phi +\frac{\lambda }{4}%
\left( \Phi ^{\dag }\Phi \right) ^{2}  \notag \\
&&+\frac{\beta }{2}\left( \Phi ^{\dag }\Phi \right) \left\langle \left(
D_{\mu }\Sigma \right) ^{\dag }D^{\mu }\Sigma \right\rangle +i\overline{\psi 
}_{iL}\gamma ^{\mu }D_{\mu }\psi _{iL}+i\overline{\psi }_{iR}\gamma ^{\mu
}D_{\mu }\psi _{iR}+y_{ij}\overline{\psi }_{iL}\Sigma \Phi \psi _{iR}+%
\widetilde{y}_{ij}\overline{\psi }_{iL}\Sigma \Phi^c \psi
_{iR}+h.c.  \label{Modellagrangian}
\end{eqnarray}%

\begin{table}[h]
    \begin{tabular}{ | c | c | c | c |}
    \hline
     $\text{Fields}$& $SU(2)_1$ & $SU(2)_2$ & $U(1)_Y$ \\ \hline
     \hline
    $\Sigma$ & $\textbf{2}$ & $\bar{\textbf{2}}$ & 0  \\
    $\Phi$ & 1 & $\textbf{2}$ & 1/2 \\
    $Q_L^i$ & $\textbf{2}$ & 1 & 1/6 \\
    $U_R^i$ & 1 & 1 & 2/3 \\ 
    $D_R^i$ & 1 & 1 & -1/3 \\ 
    $L_L^i$ & $\textbf{2}$ & 1 & -1/2 \\
    $e_R^i$ & 1 & 1 & -1 \\ 
    $N_R^i$ & 1 & 1 & 0 \\     
    \hline
    \end{tabular}
    \caption{Field charge assignments under the full gauge group $SU(2)_1\times SU(2)_2 \times U(1)_Y$. The $i$ index runs from 1 to 3.}
    \label{tabla1}
\end{table}

Here $F^{{(1)}\mu \nu }$, $F^{(2)\mu \nu }$ and $B^{\mu \nu }$ are the gauge
field tensors, the brackets $\left\langle {}\right\rangle $ denote the trace
in the corresponding group indices, $f_{\Sigma }$ is the analog of a decay
constant for the extra would-be Goldstones expressed non-linearly in the
field $\Sigma $ (these are absorbed as the longitudinal components of the
heavy vector composites), while $\lambda $ and $\mu $ are the SM parameters
of the Higgs potential, and $\beta $ is the coefficient of a mixing term
allowed by the symmetry. The value of $\beta$ is not easy to isolate from other parameters in observable quantities, so for the sake of simplicity from now on we fix $\beta=2$. Finally, the covariant derivatives are: 
\begin{eqnarray}
D_{\mu }\Sigma &=&\partial _{\mu }\Sigma -ig_{1}A^{(1)}_{\mu }\Sigma
+ig_{2}\Sigma A^{(2)}_{\mu },\hspace{0.7cm}\hspace{0.7cm}D_{\mu }\Phi =\partial
_{\mu }\Phi -ig_{2}A^{(2)}_{\mu }\Phi -i\frac{g^{\prime }}{2}B_{\mu }\Phi , 
\notag \\
D_{\mu }\psi _{iL} &=&\partial _{\mu }\psi _{iL}-ig_{1}A^{(1)}_{\mu }\psi
_{iL}-ig^{\prime }Y_{f_{iL}}B_{\mu }\psi _{iL},\hspace{0.7cm}\hspace{0.7cm}%
D_{\mu }\psi _{iR}=\partial _{\mu }\psi _{iR}-ig^{\prime }Y_{f_{iR}}B_{\mu
}\psi _{iR},
\label{CovDerivative}
\end{eqnarray}
where $A^{(n)}_{\mu }=\frac{1}{2} \tau^{a}A^{(n)a}_{\mu }$ ($n=1,2$) are
the Hermitian gauge field matrices corresponding to the $SU(2)
_{n} $ gauge fields $A^{(n)}_{\mu }$, $n=1,2$ respectively.

Notice that $\Phi $ is coupled to $A^{(2)}_{\mu}$ but not to $A^{(1)}_{\mu}$, and so it is more strongly coupled to the heavy vectors than to the SM gauge fields. In addition, left
handed SM fermionic fields will couple mainly to SM gauge fields, which are
primarly contained in $SU(2) _{1}$. The scalar doublet will
correspond to the SM Higgs field, which can be expressed as usual by: 

\begin{equation}
\Phi =\left( 
\begin{array}{c}
G^{+} \\ 
\frac{1}{\sqrt{2}}\left( v+h+iG^{0}\right)%
\end{array}%
\right) ,
\label{HiggsDoublet}
\end{equation}
where the field $h$ is the Higgs boson, while $G^{\pm }$ and $G^{0}$ are the would-be Goldstones that will be absorbed after EWSB. 
The spontaneous breaking of the extra gauge symmetry can be formulated by taking  $\Sigma =1$ (in the unitary gauge). The Lagrangian then takes the following form:

\begin{eqnarray}
\tciLaplace &=&\tciLaplace _{gauge}
+
\frac{{f_{\Sigma }}^{2}g^{2}}{4\sin^{2}2\vartheta }V^{(2)}_{\mu }V^{(2)\mu }
+
\left( D_{\mu }\Phi \right) ^{\dag}D^{\mu }\Phi 
-
\mu ^{2}\Phi ^{\dag }\Phi 
+
\frac{\lambda }{4}\left( \Phi^{\dag }\Phi \right) ^{2}
+
\frac{\beta g^{2}}{\sin ^{2}2\vartheta }\left( \Phi^{\dag }\Phi \right) V^{(2)}_{\mu }V^{(2)\mu }  
\notag \\
&&+
i\overline{\psi }_{iL}\gamma ^{\mu }D_{\mu }\psi _{iL}+i\overline{\psi }%
_{iR}\gamma ^{\mu }D_{\mu }\psi _{iR}+y_{ij}\overline{\psi }_{iL} \Phi
\psi _{iR}+\widetilde{y}_{ij}\overline{\psi }_{iL}\Phi^c%
\psi _{iR}+h.c,  \label{LU}
\end{eqnarray}%
where the covariant derivate is now rewritten as follows: 
\begin{equation}
D_{\mu }\Phi =\partial _{\mu }\Phi -ig\frac{\tau ^{a}}{2}{V}^{(1)}_{\mu , a }\Phi
-i \tilde g \frac{\tau ^{a}}{2}V^{(2)}_{\mu , a }\Phi -i\frac{g^{\prime }}{2}%
B_{\mu }\Phi ,  \label{CD}
\end{equation}%
with the vector fields given by: 
\begin{equation}
V^{(1)}_{\mu} = \cos \vartheta A^{(1)}_{\mu } + \sin \vartheta A^{(2)}_{\mu }, 
\hspace{0.7cm}
V^{(2)}_{\mu } = -\sin \vartheta   A^{(1)}_{\mu } +  \cos \vartheta A^{(2)}_{\mu },
\hspace{0.7cm}
\tan \vartheta =\frac{g_{1}}{g_{2}}, 
\end{equation}
and the couplings:
\begin{equation}
g = \frac{g_1 g_2}{\sqrt{g_1^2+g_2^2}}, \hspace{0.7cm}
\tilde g = \frac{g_2^2}{\sqrt{g_1^2+g_2^2}} .  \label{Relations}
\end{equation}

At this stage, the fields $V^{(1)}_{\mu}$ remain massless but $V^{(2)}_{\mu}$ acquire mass proportional to $f_\Sigma$, as shown in Eq.~\ref{LU}.
When the Higgs boson acquires a vacuum expectation value $\left\langle \Phi
\right\rangle =\frac{1}{\sqrt{2}}\left( 0,v\right) ^{T}$, from Eq. (\ref{LU}%
) it follows that the squared mass matrices for the neutral and charged
gauge bosons are given by: 
\begin{equation}
M_{N}^{2}=\frac{v^{2}}{4}
\left( 
\begin{array}{ccc}
g^{2} & -gg^{\prime } & g{\tilde g} \\ 
-gg^{\prime } & g^{\prime 2} & -g^{\prime }{\tilde g} \\ 
g{\tilde g} & -g^{\prime }{\tilde g} & {\tilde g}^{2}k^{2}%
\end{array}
\right) ,
\hspace{0.7cm}\hspace{0.7cm}
M_{C}^{2}=\frac{v^{2}}{4}
\left( 
\begin{array}{cc}
g^{2} & g{\tilde g} \\ 
g{\tilde g} & k^{2}{\tilde g}^{2}%
\end{array}
\right) ,
\hspace{1cm}\textrm{with}\ \ \ k^{2}=1+\frac{{2f_{\Sigma }}^{2}+\beta v^{2}}{v^{2}\cos ^{4}\vartheta }.  \label{Gaugebosons}
\end{equation}

The masses of the gauge bosons are given by diagonalization of these mass matrices:

\begin{eqnarray}
M_{A} &=&0,  \notag \\
M_{Z} &=&\frac{v}{2\sqrt{2}}\sqrt{g^{2}+g^{\prime 2}+{\tilde g}^{2}k^{2}-%
\sqrt{{\tilde g}^{4}k^{4}+\left( g^{2}+g^{\prime 2}\right) \left[
g^{2}+g^{\prime 2}+2\left( 2-k^{2}\right) {\tilde g}^{2}\right] }},  \notag
\\
M_{\rho ^{0}} &=&\frac{v}{2\sqrt{2}}\sqrt{g^{2}+g^{\prime 2}+\tilde g^{2}k^{2}+\sqrt{{\tilde g}^{4}k^{4}+\left( g^{2}+g^{\prime 2}\right) \left[
g^{2}+g^{\prime 2}+2\left( 2-k^{2}\right) {\tilde g}^{2}\right] }},  \notag
\\
M_{W^\pm} &=&\frac{v}{2\sqrt{2}}\sqrt{k^{2}{\tilde g}^{2}+\allowbreak g^{2}-%
\sqrt{\left( g^{2}-k^{2}{\tilde g}^{2}\right) ^{2}+4g^{2}{\tilde g}^{2}}}, 
\notag \\
M_{\rho ^{\pm }} &=&\frac{v}{2\sqrt{2}}\sqrt{k^{2}{\tilde g}^{2}+\allowbreak
g^{2}+\sqrt{\left( g^{2}-k^{2}{\tilde g}^{2}\right) ^{2}+4g^{2}{\tilde g}^{2}%
}},  \label{massesofgaugebosons}
\end{eqnarray}

and the physical neutral and charged gauge bosons are given by:%
\begin{equation}
\left( 
\begin{array}{c}
A_{\mu } \\ 
Z_{\mu } \\ 
\rho _{\mu }^{0}%
\end{array}%
\right) =\left( 
\begin{array}{ccc}
\cos \theta_W & \sin \theta_W & 0 \\ 
-\cos \gamma \sin \theta_W & \cos \gamma \cos \theta_W & -\sin \gamma
\\ 
-\sin \gamma \sin \theta_W & \cos \theta_W\sin \gamma & \cos \gamma%
\end{array}%
\right) \left( 
\begin{array}{c}
B_{\mu }^{0} \\ 
V_{\mu , 3}^{(1)} \\ 
V_{\mu , 3}^{(2)}%
\end{array}%
\right) ,\hspace{0.7cm}\left( 
\begin{array}{c}
W_{\mu }^{\pm } \\ 
\rho _{\mu }^{\pm }%
\end{array}%
\right) =\left( 
\begin{array}{cc}
\cos \kappa & -\sin \kappa \\ 
\sin \kappa & \cos \kappa%
\end{array}%
\right) \left( 
\begin{array}{c}
V_{\mu }^{(1) \pm } \\ 
V_{\mu }^{(2) \pm }%
\end{array}%
\right) ,
\end{equation}

where, besides the standard $\theta_W$,  the additional mixing angles  $\kappa$ and $\gamma$ are:
\begin{equation}
\tan 2\kappa =
\frac{
    \sqrt{
\left( 4M_{\rho ^{\pm }}^{2}-g^{2}v^{2}\right)
\left( g^{2}v^{2}-4M_{W}^{2}\right) 
    }
}
{
2\left( M_{\rho ^{\pm}}^{2}+M_{W}^{2}\right) -g^{2}v^{2}
},
\hspace{0.7cm}\hspace{0.7cm}
\tan 2\gamma =
\frac
{2 \tilde g  \sqrt{g^2+g^{\prime 2} }  }  {{\tilde g}^{2}k^{2}-g^{2}-g^{\prime 2}}.
\end{equation}

At this stage the electroweak symmetry is finally broken and the only remaining massless vector boson  is the photon field $A_\mu$. 

\section{Constraints from Higgs decay into two photons}

\label{Higgsdiphotonrate}

\begin{figure}[tbh]
\resizebox{0.7\textwidth}{!}
{\includegraphics{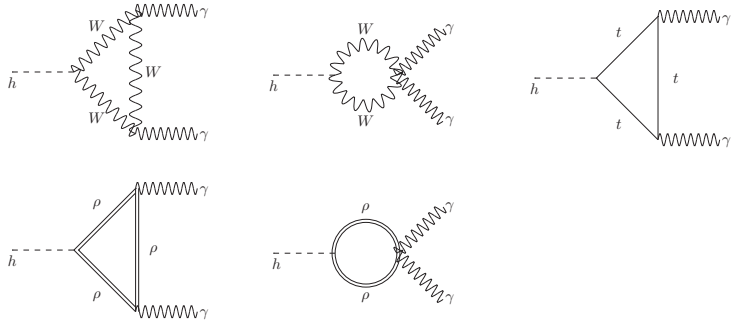}}\vspace{-10cm}
\caption{One loop Feynman diagrams in the Unitary Gauge contributing to the $%
h\rightarrow \protect\gamma \protect\gamma $ decay.}
\label{figura1}
\end{figure}

In the Standard Model, the $h\rightarrow \gamma \gamma $ decay is dominated
by $W$ loop diagrams which can interfere destructively with the subdominant
top quark loop. In our strongly coupled model, the $h\rightarrow
\gamma\gamma $ decay receives additional contributions from loops with
charged $\rho _{\mu }^{\pm }$, as shown in Fig.~\ref{figura1}. The
explicit form for the $h\rightarrow \gamma \gamma $ decay rate is: 
\begin{equation}
\Gamma \left( h\rightarrow \gamma \gamma \right) =\frac{\alpha
_{em}^{2}m_{h}^{3}}{256\pi ^{3}v^{2}}\left\vert
\sum_{f}N_{c}Q_{f}^{2}F_{1/2}\left( x_{f}\right) +a_{hWW}F_{1}\left(
x_{W}\right) +a_{h\rho ^{+}\rho ^{-}}F_{1}\left( x_{\rho }\right)
\right\vert ^{2},
\end{equation}%
where:

\begin{eqnarray}
a_{hW^{+}W^{-}} &=&\frac{vg^{2}}{2}\left\{ \left( \cos \kappa -\cot \vartheta
\sin \kappa \right) ^{2}+\frac{\beta \sin ^{2}\kappa }{\sin ^{2}\vartheta \cos
^{2}\vartheta }\right\} \frac{v}{2M_{W}^{2}},\hspace{0.3cm}  \label{coupling1}
\\
a_{h\rho ^{+}\rho ^{-}} &=&\frac{vg^{2}}{2}\left\{ \left( \sin\kappa+\cot
\vartheta \cos\kappa \right) ^{2}+\frac{\beta \cos ^{2}\kappa }{\sin ^{2}\vartheta
\cos ^{2}\vartheta }\right\} \frac{v}{2M_{\rho ^{\pm }}^{2}}.
\end{eqnarray}

Here $x_{i}$ are the mass ratios $x_{i}=m_{h}^{2}/4M_{i}^{2}$, with $%
M_{i}=m_{f},M_{W}$ and $M_{\rho }$, respectively, $\alpha _{em}$ is the fine
structure constant, $N_{C}$ is the color factor ($N_{C}=1$ for leptons, $%
N_{C}=3$ for quarks), and $Q_{f}$ is the electric charge of the fermion in
the loop. From the fermion loop contributions we will keep only the dominant
term, which is the one involving the top quark.

The dimensionless loop factors $F_{1/2}\left( x\right) $ and $F_{1}\left(
x\right) $ (for particles of spin $1/2$ and 1 in the loop, respectively) are 
\cite%
{Shifman:1979eb,Gavela:1981ri,Kalyniak:1985ct,Gunion:1989we,Spira:1997dg,Djouadi:2005gj,Marciano:2011gm,Wang:2012gm}%
: 
\begin{equation}
F_{1/2}\left( x \right) =2\left[ x+\left( x-1\right) f\left( x\right) \right]
x^{-2},
\end{equation}%
\begin{equation}
F_{1}\left( x \right) =-\left[ 2x^{2}+3x+3\left( 2x-1\right) f\left(
x\right) \right] x^{-2},  \label{F}
\end{equation}%
with 
\begin{equation}
f\left( x\right) =%
\begin{cases}
\arcsin ^{2}\sqrt{x},\hspace{0.5cm}\mathit{for}\hspace{0.2cm}x\leq 1 \\ 
-\frac{1}{4}\left[ \ln \left( \frac{1+\sqrt{1-x^{-1}}}{1-\sqrt{1-x^{-1}}}%
\right) -i\pi \right] ^{2},\hspace{0.5cm}\mathit{for}\hspace{0.2cm}x>1.%
\end{cases}%
\end{equation}

In what follows, we want to determine the range of values for the mass $%
M_{\rho }$ of the heavy vector resonances and the mixing angle $\vartheta$, consistent with the
Higgs diphoton signal strength measured by the ATLAS and CMS collaborations
at the LHC. 
To this end, we introduce the ratio $R_{\gamma \gamma }$, which corresponds
to the Higgs diphoton signal strength that normalises the $\gamma \gamma $
signal predicted by our model relative to that of the SM: 
\begin{equation}
R_{\gamma \gamma }=\frac{\sigma \left( pp\rightarrow h\right) \Gamma \left(
h\rightarrow \gamma \gamma \right) }{\sigma \left( pp\rightarrow h\right)
_{SM}\Gamma \left( h\rightarrow \gamma \gamma \right) _{SM}}\simeq \frac{%
\Gamma \left( h\rightarrow \gamma \gamma \right) }{\Gamma \left(
h\rightarrow \gamma \gamma \right) _{SM}}.  \notag
\end{equation}

This normalization for $h\rightarrow \gamma \gamma $ was also done in Refs.~%
\cite{Wang:2012gm,Campos:2014zaa,Hernandez:2015dga}. Here we have used the
fact that in our model, single Higgs production is also dominated by gluon
fusion as in the Standard Model.

\begin{figure}[tbh]
\resizebox{0.5\textwidth}{!}{
\includegraphics{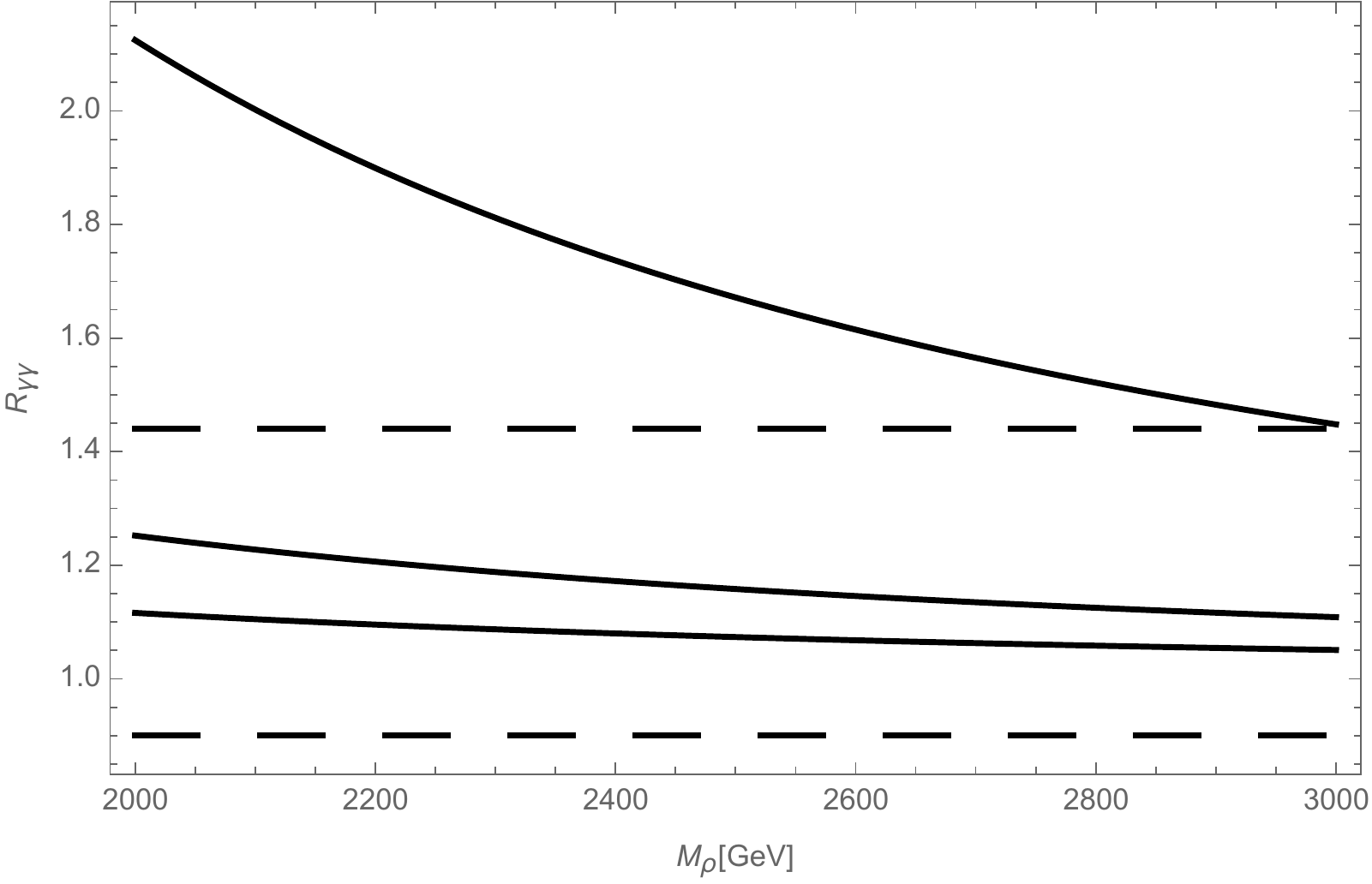}
} 
\caption{The ratio $R_{\protect\gamma \protect\gamma }$ as a function of $M_{%
\protect\rho }$ for several values of $\tan\protect\vartheta$. 
The solid curves from top to bottom correspond to $\tan\protect\vartheta=0.1,\ 0.2$ and $0.3$. The horizontal dashed lines are the minimum and maximum values
of the ratio $R_{\protect\gamma \protect\gamma }$ inside the 1$\protect%
\sigma $ experimentally allowed range by CMS and ATLAS, namely  $%
1.14_{-0.23}^{+0.26}$ and $1.17\pm 0.27$, respectively \protect\cite%
{Khachatryan:2014ira,Aad:2014eha}.}
\label{fig2}
\end{figure}

\begin{figure}[tbh]
\resizebox{0.35\textwidth}{!}
{\includegraphics{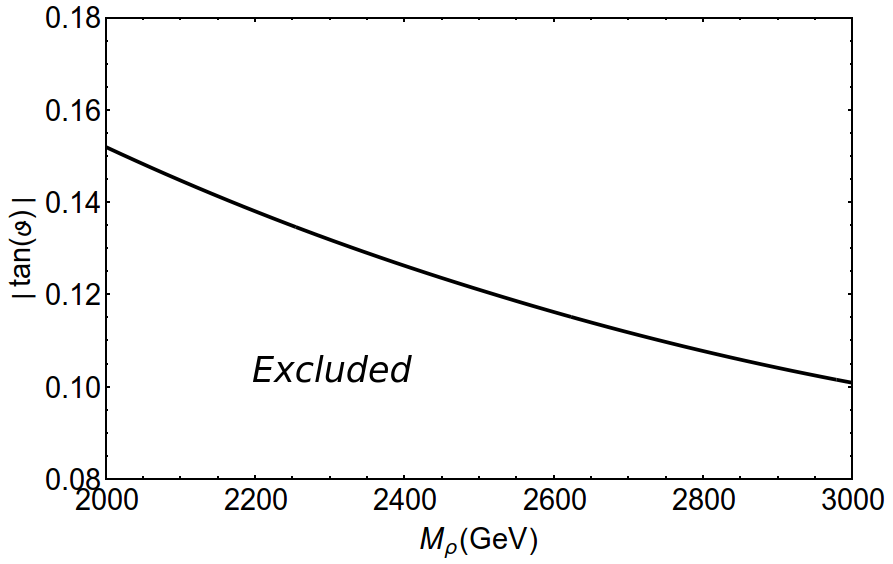}}
\caption{Lower bound on $\tan\protect\vartheta$ vs. $M_\rho$, 
according to the constraint imposed by the Higgs diphoton decay rate $h\to \gamma \gamma$ at the LHC.}
\label{fig3}
\end{figure}

Fig.~\ref{fig2} shows the sensitivity of the ratio $R_{\gamma \gamma }$ under
variations of $M_{\rho }$ for several values of $\tan\protect\vartheta$. The curves from top to bottom correspond to $\tan\protect%
\vartheta=0.1,0.2,0.3$.
The ratio $R_{\gamma \gamma }$ decreases slowly when the heavy vector masses
are increased. 

As shown, our model successfully accommodates the current Higgs diphoton decay rate
constraints. 

A more exhaustive study of the allowed values of $\vartheta$ for different $M_\rho$ is shown in 
Fig.~\ref{fig3}. The observed Higgs diphoton decay rate at the LHC excludes the white region below the curve in the figure, 
corresponding to too small values of $\vartheta$: for such small $\vartheta$ values the Higgs boson would couple too 
strongly to the heavy vector resonances, increasing the Higgs diphoton decay rate beyond the observed values.
In addition, the heavy vector contribution to the
Higgs diphoton decay rate scales as $1/M_\rho^2$ due to the heavy vector
propagator and consequently, as  Fig.~\ref{fig3} shows, the tightest lower bound $\tan\vartheta \gtrsim 0.15$ is obtained at the lower en of $M_\rho$; for larger masses of the vector resonances the $\vartheta$ values are less restricted.

\section{Constraints from the T and S parameters}

\label{TandS}

The inclusion of the extra composite particles also modifies the oblique
corrections of the SM, the values of which have been extracted from high
precision experiments. Consequently, the validity of our model depends on
the condition that the extra particles do not contradict those experimental
results. These oblique corrections are parametrized in terms of the two well
known quantities $T$ and $S$. The $T$ parameter is defined as \cite%
{Peskin:1991sw,Altarelli:1990zd,Barbieri:2004qk}: 
\begin{equation}
T=\frac{\Pi _{33}\left( 0\right) -\Pi _{11}\left( 0\right) }{M_{W}^{2}\alpha
_{em}\left( m_{Z}\right) },  \label{T}
\end{equation}%
where $\Pi _{33}\left( 0\right) $ and $\Pi _{11}\left( 0\right) $ are the
vacuum polarization amplitudes at $q^{2}=0$ for the propagators of the gauge
bosons $A_{\mu , 3}^{(1)}$ and $A_{\mu ,1}^{(1) }$, respectively, which are those that couple to the external fermions in the process $e^+ e^- \to f \bar f$ \cite{Barbieri:2004qk}. 

\begin{figure*}[tbh]
\resizebox{0.95\textwidth}{!}{
\includegraphics{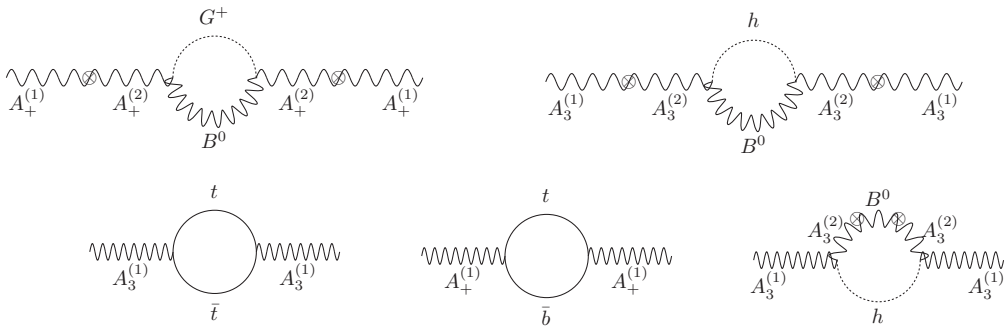}}\vspace{-16cm}
\caption{One loop Feynman diagrams contributing to the $T$ parameter. The fields are those in Eqs.~(\ref{CovDerivative}) and (\ref{HiggsDoublet}).}
\label{figT}
\end{figure*}

\begin{figure*}[tbh]
\resizebox{0.95\textwidth}{!}{\includegraphics{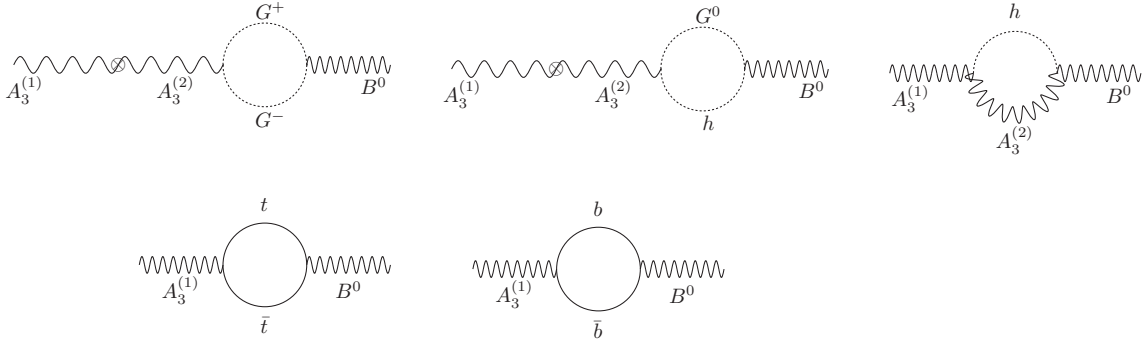}}\vspace{-16cm}
\caption{One loop Feynman diagrams contributing to the $S$ parameter. 
The fields are those in Eqs.~(\ref{CovDerivative}) and (\ref{HiggsDoublet}).}
\label{figS}
\end{figure*}
In turn, the $S$ parameter is defined as \cite%
{Peskin:1991sw,Altarelli:1990zd,Barbieri:2004qk}: 
\begin{equation}
S=\frac{4\sin ^{2}\theta_W}{\alpha _{em}\left( m_{Z}\right) }\frac{g}{%
g^{\prime }}\frac{d}{dq^{2}}\Pi _{30}\left( q^{2}\right) \biggl|_{q^{2}=0},
\end{equation}
where $\Pi _{30}\left( q^{2}\right) $ is the vacuum polarization 
for the propagator mixing of $A^{(1)}_{\mu , 3 }$ and $B_{\mu }$. 
The most
important Feynman diagrams contributing to the $T$ and $S$ parameters are
shown in Figures \ref{figT} and \ref{figS}. We computed these oblique $T$
and $S$ parameters in the Landau gauge for the SM gauge bosons and
would-be-Goldstone bosons, where the global $SU(2)_{L}\times U(1)_{Y}$
symmetry is preserved. 
We can separate 
the contributions to $T$ and $S$ from the SM and extra physics 
as $T=T_{SM}+\Delta T$ and $S=S_{SM}+\Delta S$,
where 
\begin{eqnarray}
T_{SM} &=&-\frac{3}{16\pi \cos ^{2}\theta_W}\ln \left( \frac{m_{h}^{2}}{%
m_{W}^{2}}\right) +\frac{3m_{t}^{2}}{32\pi ^{2}\alpha _{em}\left(
m_{Z}\right) v^{2}},\hspace{0.2cm}  
\notag \\
S_{SM} &=&\frac{1}{12\pi }\ln \left( \frac{m_{h}^{2}}{m_{W}^{2}}\right) +%
\frac{1}{2\pi }\left[ 3-\frac{1}{3}\ln \left( \frac{m_{t}^{2}}{m_{b}^{2}}%
\right) \right] ,
\end{eqnarray}

while $\Delta T$ and $\Delta S$ contain all the contributions involving the
extra particles. 

The dominant one-loop contribution to $\Delta T$ and $\Delta S$ in our model
are: 
\begin{eqnarray}
\Delta T=\tan^2\vartheta \ T_{SM}-\frac{3\beta ^{2}M_{W}^{4}}{16\pi \cos ^{2}\vartheta \sin ^{4}\vartheta \cos
^{2}\theta _{W}}F\left( M_{B},m_{h},M_{A^{\left( 2\right) }}\right),
\label{DeltaT}
\end{eqnarray}
\begin{equation}
\Delta S =\frac{1-\cos\vartheta}{\cos\vartheta}\ S_{SM} + \frac{2\beta M_{W}^{2}}{\pi\sin ^{2}\vartheta \cos
\vartheta}\left[ G_{1}\left( M_{A^{\left( 2\right) }},m_{h}\right) -\frac{1%
}{4M_{A^{\left( 2\right) }}^{2}}G_{2}\left( M_{A^{\left( 2\right)
}},m_{h}\right) \right]
,
\label{DeltaS}
\end{equation}
where 
\begin{eqnarray}
F\left( m_{1},m_{2},m_{3}\right)  
&=&
\frac{m_{1}^{2}}{\left( m_{1}^{2}-m_{2}^{2}\right) \left( m_{1}^{2}-m_{3}^{2}\right) ^{2}}
\ln \left( \frac{ 1+\Lambda ^{2}/m_{1}^{2}} {1 + \Lambda^2/m_3^2} \right) 
+
\frac{m_{2}^{2}}{\left( m_{2}^{2}-m_{1}^{2}\right) \left( m_{2}^{2}-m_{3}^{2}\right) ^{2}}
\ln \left( \frac{1+ \Lambda ^{2}/m_{2}^{2}}{1+\Lambda^2/m_3^2}\right)
\notag \\
&&+
\frac{m_{3}^{2}}{\left( m_{3}^{2}-m_{1}^{2}\right) \left( m_{3}^{2}-m_{2}^{2}\right) }
\left[ \frac{1}{\Lambda ^{2}+m_{3}^{2}}-\frac{1}{m_{3}^{2}}\right] ,
\end{eqnarray}
\begin{equation}
G_{1}\left( m_{1},m_{2}\right) =\int_{0}^{1}dx\ x\left(
1-x\right) \left[ \frac{\frac{1}{2}\left[ -\left( m_{1}^{2}-m_{2}^{2}\right)
x+m_{1}^{2}\right] +\Lambda ^{2}}{\left( \left[ -\left(
m_{1}^{2}-m_{2}^{2}\right) x+m_{1}^{2}\right] +\Lambda ^{2}\right) ^{2}}-%
\frac{1}{2\left[ -\left( m_{1}^{2}-m_{2}^{2}\right) x+m_{1}^{2}\right] }%
\right] ,
\end{equation}
\begin{eqnarray}
G_{2}\left( m_{1},m_{2}\right) &=&\int_{0}^{1}dx \ x\left(
1-x\right) \left[ \ln \left( \frac{\Lambda ^{2}+\left[ -\left(
m_{1}^{2}-m_{2}^{2}\right) x+m_{1}^{2}\right] }{\left[ -\left(
m_{1}^{2}-m_{2}^{2}\right) x+m_{1}^{2}\right] }\right) -\frac{3}{2}\right]\notag \\
&&+\int_{0}^{1}dx\  x\left( 1-x\right) \frac{4\left[
-\left( m_{1}^{2}-m_{2}^{2}\right) x+m_{1}^{2}\right] \Lambda ^{2}+3\left[
-\left( m_{1}^{2}-m_{2}^{2}\right) x+m_{1}^{2}\right] ^{2}}{2\left\{ \Lambda
^{2}+\left[ -\left( m_{1}^{2}-m_{2}^{2}\right) x+m_{1}^{2}\right] \right\}
^{2}} ,
\end{eqnarray}
\begin{equation}
M_{A^{\left( 2\right) }}=\frac{g_{2}}{\sqrt{2}}\sqrt{f_{\Sigma }^{2}+\frac{%
\beta }{2}v^{2}}=\frac{4\cos ^{2}\vartheta }{\sin \vartheta }\left( \frac{%
M_{W}}{v}\right) ^{2}M_{\rho } .
\end{equation}

It is worth mentioning that we do not consider the tree level contribution of the heavy
vectors to the $\Delta T$ and $\Delta S$ parameters, since they are of the form $\sim
v^{2}/M_{\rho }^{2}$, which are subleading compared to the loop contributions.

As a result, the experimental constraints on the $T$ and $S$ parameters \cite{Baak:2011ze} impose an upper bound on our mixing parameter $\tan\vartheta \lesssim 0.47$, for heavy vector masses from $2$ TeV up to $3$ TeV.

\section{Production and decays of the heavy vectors}

\label{Decaychannels}

The current important period of LHC exploration of the Higgs properties and
discovery of heavier particles may provide crucial steps to unravel the
electroweak symmetry breaking mechanism. Consequently, we complement our
work by studying the production and decay channels of the heavy vector
resonances which are relevant for the
LHC.

\begin{figure}[tbh]
	
		\includegraphics[scale=0.45]{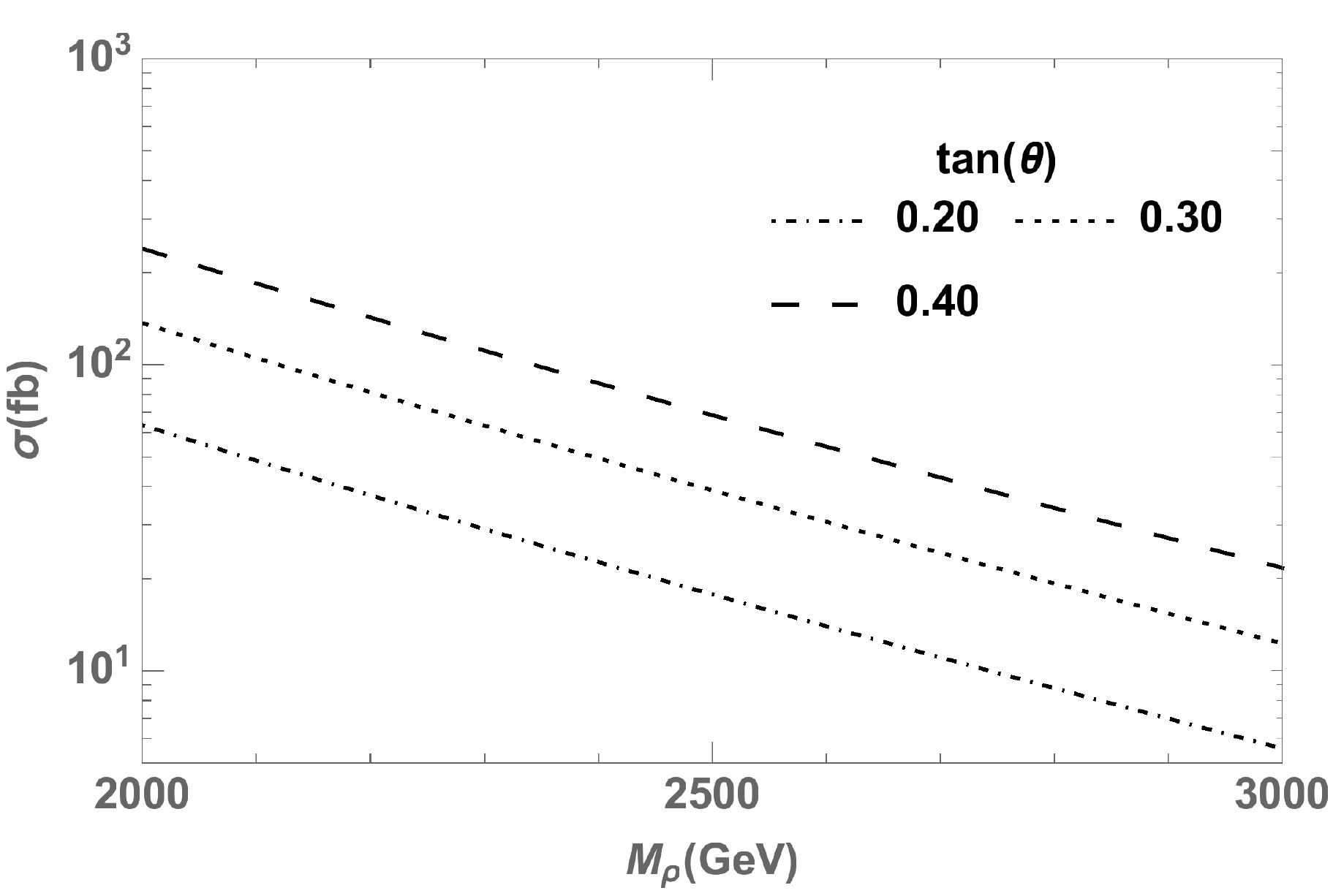} 
	
		\caption {Heavy vector production cross section $\sigma (pp\to \rho )$ vs. $M_\rho$, for $\tan\vartheta$ = 0.2, 0.3 and 0.4.	}
\label{VH3}
\end{figure}

At a hadron collider like the LHC, the most important production channel for 
the heavy vector resonance is quark--anti-quark annihilation. In our construction,
the coupling of heavy vectors to quarks goes through a term which has its origin in the mixing between the gauge fields $A^{(1)}_{\mu }$ and $A^{(2)}_{\mu }$. Consequently, the $\rho$ production amplitude is proportional to $\tan \vartheta$ which acts as a suppression factor. The influence of  $\tan \vartheta$ can be seen in Figure \ref{VH3} where we show the $\rho$ production cross section, computed with CalcHEP \cite{Belyaev:2012qa}, for different values of $M_{\rho}$ and $\tan \vartheta$.

\begin{figure*}[h!]
\resizebox{16cm}{5.5cm}
{\includegraphics{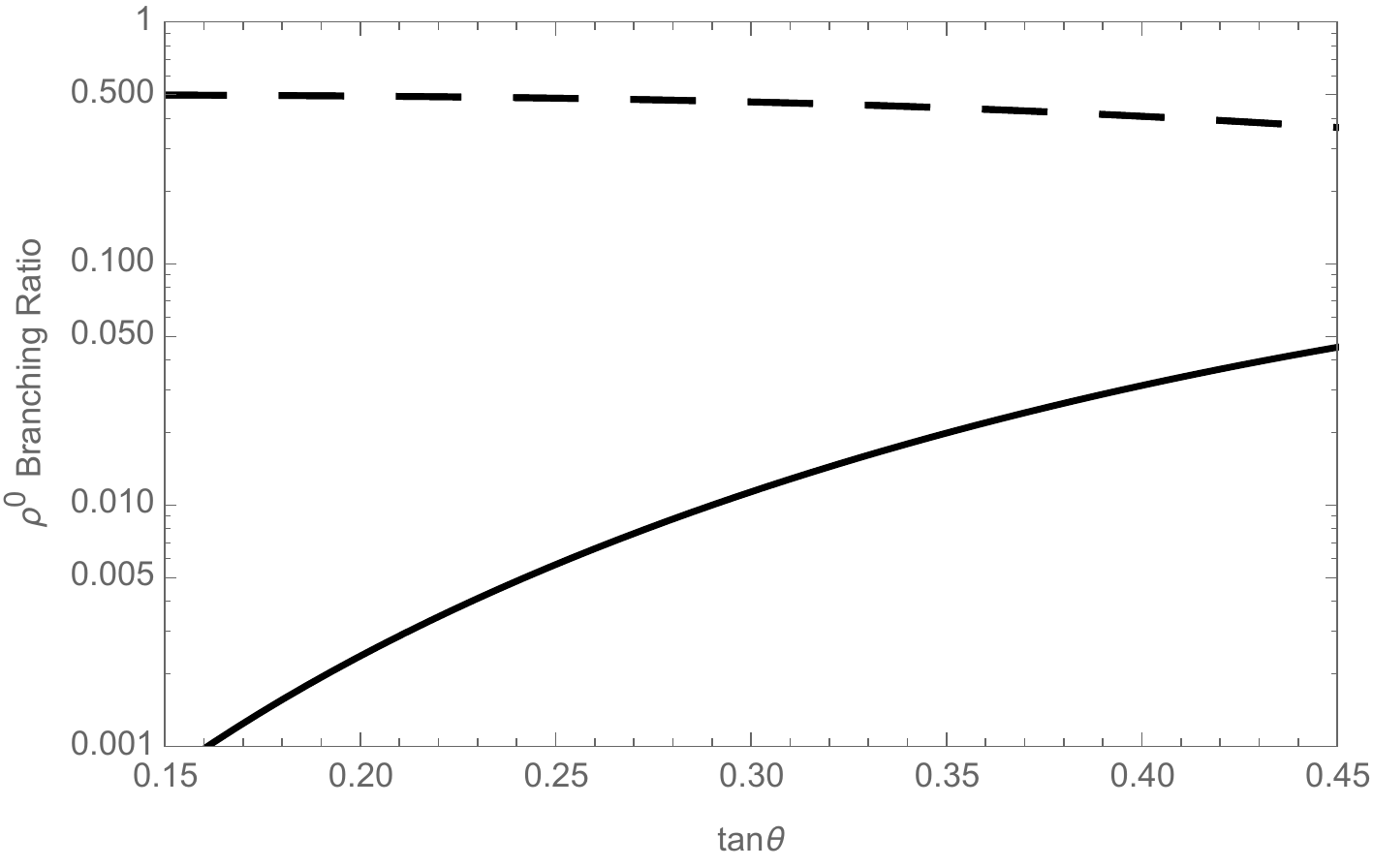} \includegraphics{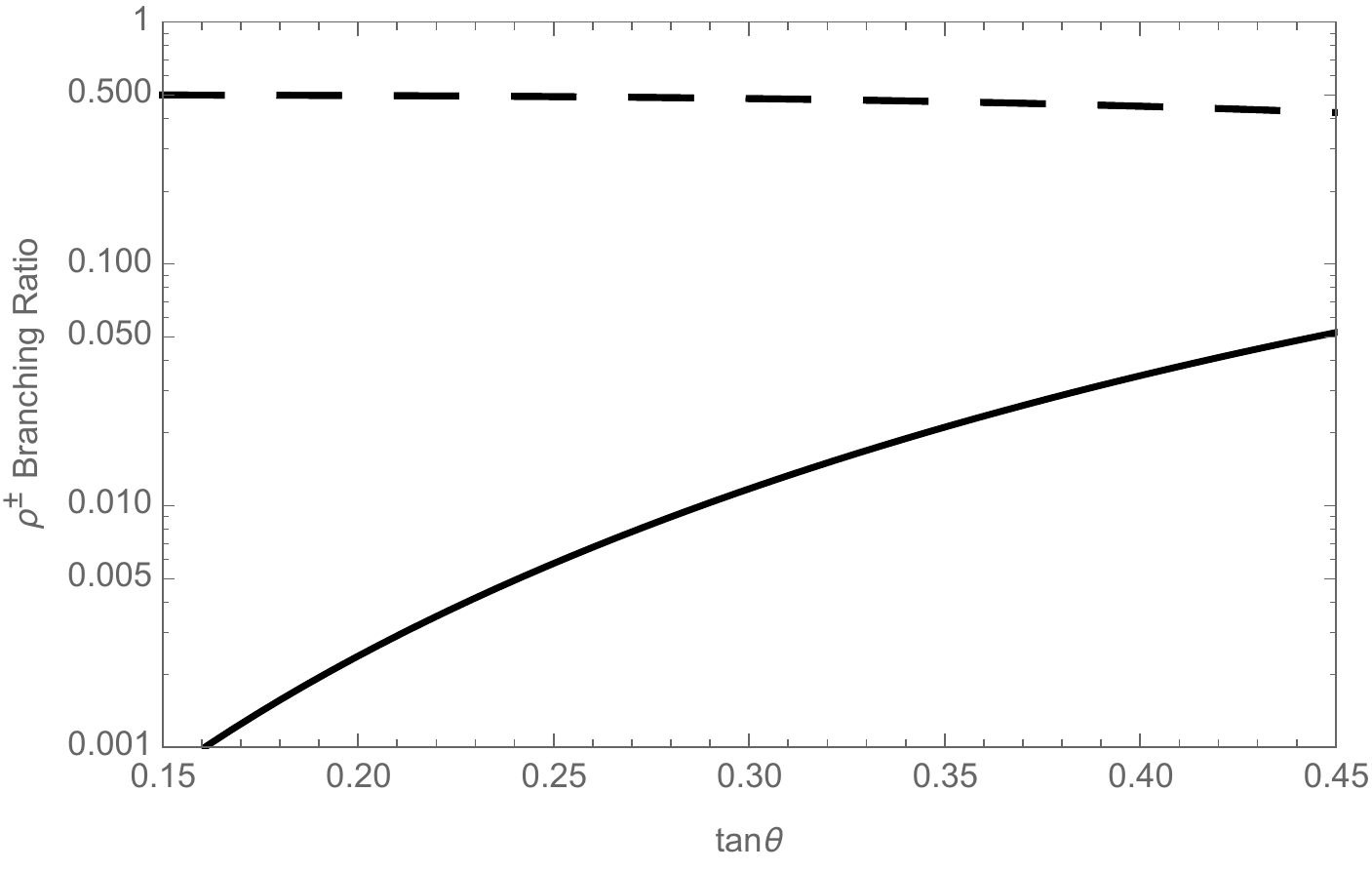}} 
\begin{tabular}{ccc}
 (a) & \hspace{8cm} & (b)%
\end{tabular}%
\caption{Branching ratios of neutral and charged heavy vector decays  vs.  $\tan\vartheta$:  (a) neutral vector decays:
$Br\left(\rho ^{0}\rightarrow q \overline{q}\right)$, 
$q= u, d, s,\ldots$ (solid)  and  $Br\left(\protect\rho^{0}\rightarrow
W^{+}W^{-}\right)=Br\left( \protect\rho^{0}\rightarrow Zh\right)$ (dashed);   
(b)  charged vector decays:
 $Br\left(\rho^{+}\rightarrow u\overline{d}\right)=
Br\left(\rho^{-}\rightarrow d\overline{u}\right)$ (solid)  and  
$Br\left(\protect\rho^{\pm}\rightarrow W^{\pm}Z\right)= Br\left( \protect\rho^{\pm }\rightarrow W^{\pm }h\right)$ (dashed). }
\label{BR}
\end{figure*}

 Additionally, we compute the two-body decay rates of the heavy vectors.
These rates, up to corrections of order $m_{h}^{2}/M_{\rho }^{2}$ and $%
M_{W}^{2}/M_{\rho }^{2}$ are: 
\begin{eqnarray}
\Gamma \left( \rho ^{0}\rightarrow q\overline{q}\right) &\simeq &\frac{%
3g^{2}\tan ^{2}\vartheta }{96\pi }M_{\rho },  \notag \\
\Gamma \left( \rho ^{+}\rightarrow u_{i}\overline{d}_{j}\right) &=&\Gamma
\left( \rho ^{-}\rightarrow \overline{u}_{i}d_{j}\right) \simeq \frac{%
3g^{2}\tan ^{2}\vartheta }{96\pi }\left\vert V_{ij}\right\vert ^{2}M_{\rho }, 
\notag \\
\Gamma \left( \rho ^{\pm }\rightarrow W^{\pm }h\right) &=&\Gamma \left( \rho
^{0}\rightarrow Zh\right) \simeq \frac{g^{2}\cot ^{2}\vartheta }{96\pi }M_{\rho
},  \notag \\
\Gamma \left( \rho ^{0}\rightarrow W^{+}W^{-}\right) &=&\Gamma \left( \rho
^{\pm }\rightarrow W^{\pm }Z\right) \simeq \frac{g^{2}\cot ^{2}\vartheta }{%
96\pi }M_{\rho }.
\end{eqnarray}

Fig.~\ref{BR} displays the branching ratios of the neutral (a) and charged (b) heavy vectors to a quark-antiquark pair
and to a SM-like Higgs in association with a SM gauge boson, as a function
of $\tan \vartheta $. This angle controls the strength of the coupling of
the heavy vector resonances with fermions. 
Clearly the largest decay rates of the heavy vectors are into a pair of SM Gauge bosons 
as well as into a SM-like
Higgs and SM gauge boson, for all values of $\tan \vartheta $. 
The decays into quark-antiquark pairs are much smaller in the relevant region of parameter space. This is a direct
consequence of the gauge structure of the model and the representations of
the fermions and the Higgs doublet under the full gauge symmetry group.


\section{Bounds from LHC searches}
\label{LHCsearches}

The ATLAS and CMS collaborations have performed several searches for heavy resonances decaying into different final states \cite{ATLAS:2017wce,Khachatryan:2016qkc,Aaboud:2017efa,TheATLAScollaboration:2016wfb,ATLAS:2016lvi}. 
These searches are based on upper limits in the resonant cross section for different heavy vector particles. We use those limits to set restrictions on the model parameter space thus complementing the diphoton and the electroweak precision test constraints described above. 
As stated at the end of Sections \ref{Higgsdiphotonrate} and \ref{TandS}, the allowed mixing parameter 
$\tan\vartheta$ is restricted to the range $0.15 \lesssim \tan\vartheta \lesssim 0.47$. In what follows, we will use as benchmark points the values $\tan\vartheta$ = 0.15, 0.20, 0.30 and 0.47.

\begin{figure}[h!]
\resizebox{17cm}{6cm}
{\includegraphics[scale=1.54]{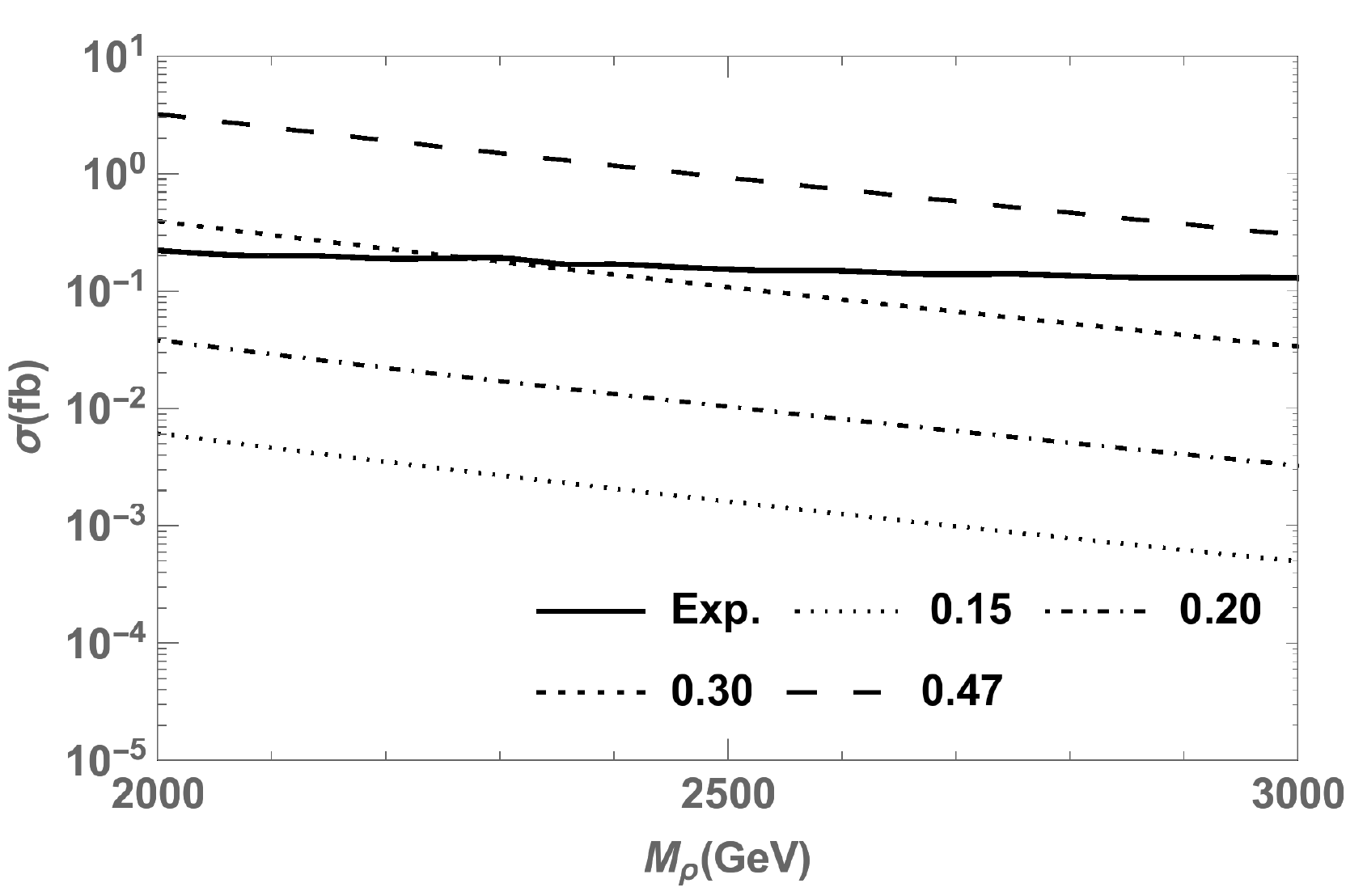}\includegraphics[scale=1.5]{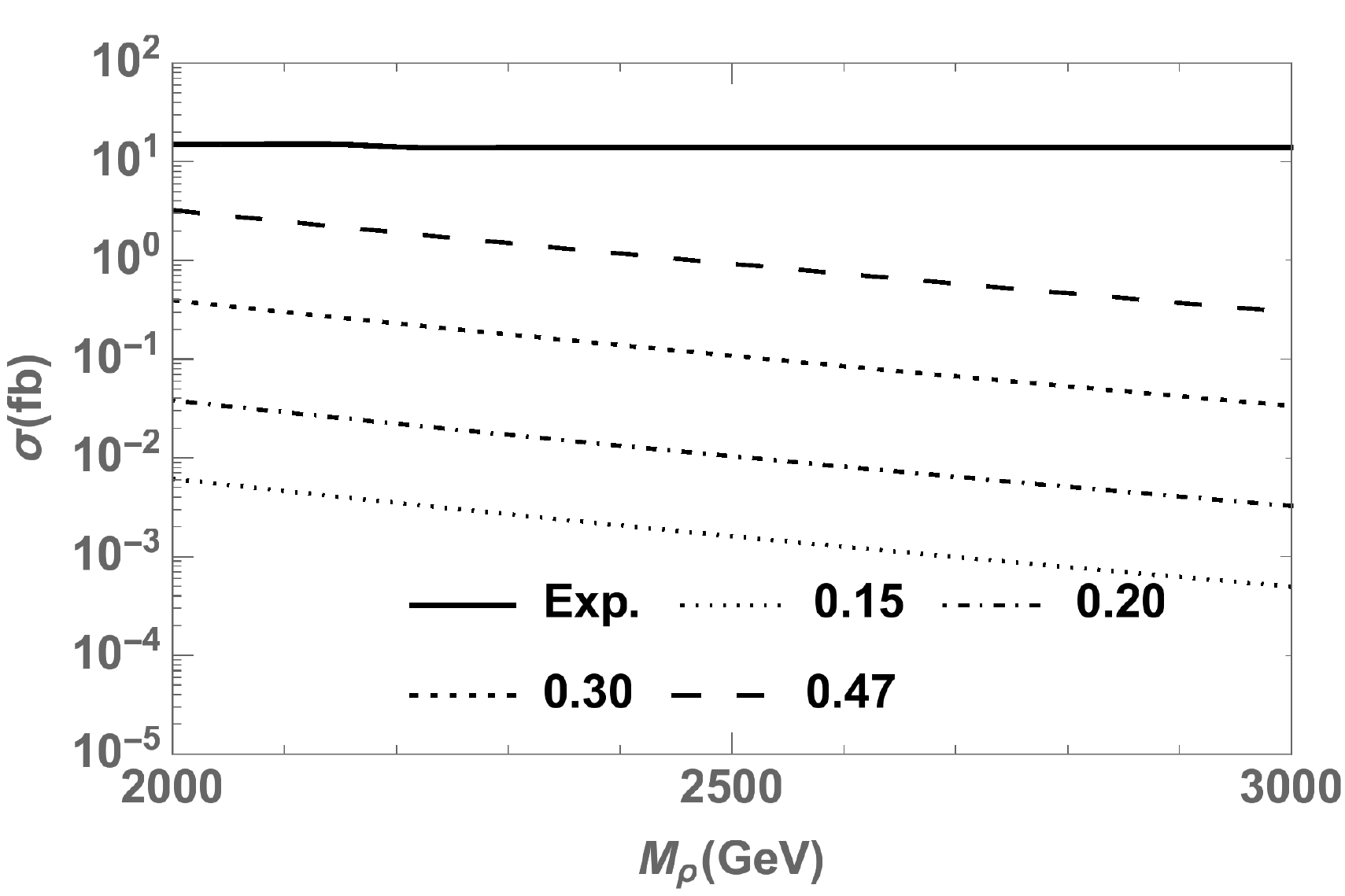}} 
\caption{Left:  predicted $p p \to \rho^0\to l^+l^-$ resonant production at $\sqrt{s}$ = 13 TeV, for the combined channel $ee$ and $\mu\mu$, as a function of $M_\rho$, for different values of $\tan\vartheta$; the solid line is the 95\% C.L. upper limit obtained  by ATLAS \cite{ATLAS:2017wce}. 
Right: \emph{idem}, but for the $pp \to \rho^0 \to \tau^+\tau^-$ channel; the solid line is the 95\% C.L. upper limit obtained by CMS \cite{Khachatryan:2016qkc}.}
\label{ULlep}
\end{figure}

\begin{figure}[h!]
\resizebox{9cm}{6cm}
{\includegraphics[scale=1.5]{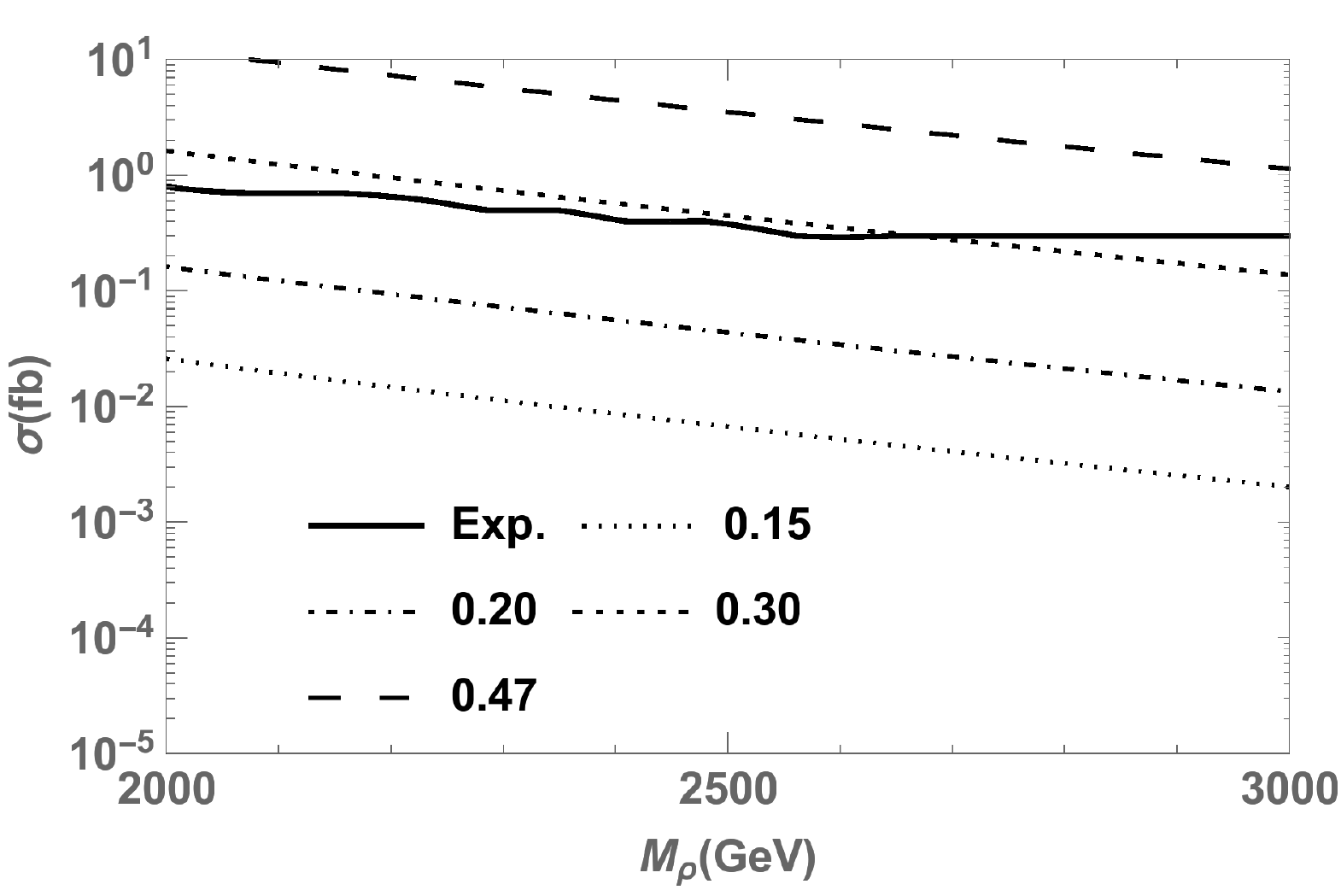}} 
\caption{
 Predicted $p p \to \rho^\pm\to l \nu$ resonant production at $\sqrt{s}$ = 13 TeV, for $l= e, \mu$, as a function of $M_\rho$, for different values of $\tan\vartheta$; the solid line is the 95\% C.L. upper limit obtained  by ATLAS \cite{ATLAS:2017wce}.} 
\label{ULlv}
\end{figure}

\begin{figure}[h!]
\resizebox{17cm}{6cm}
{\includegraphics{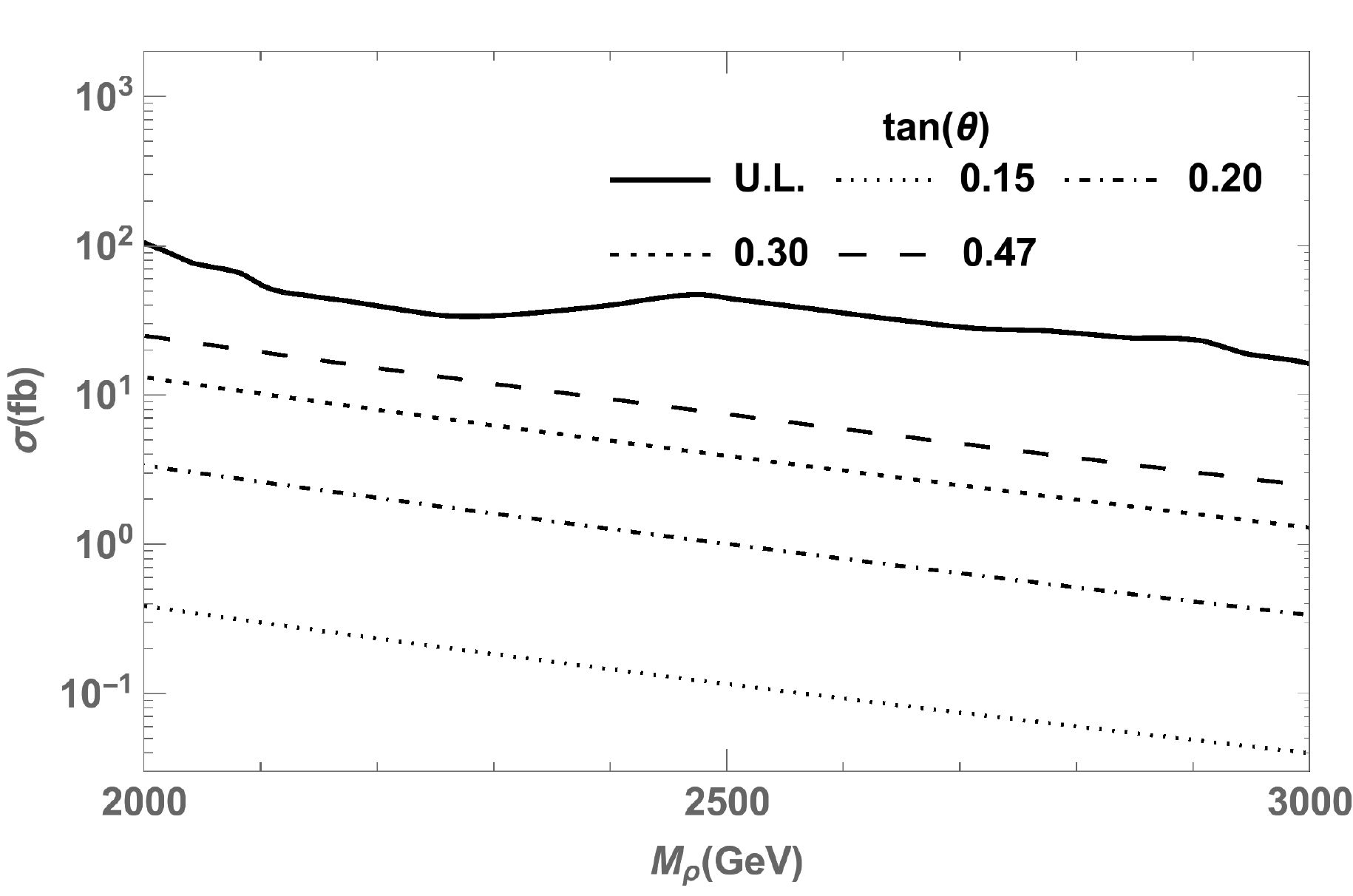}\includegraphics{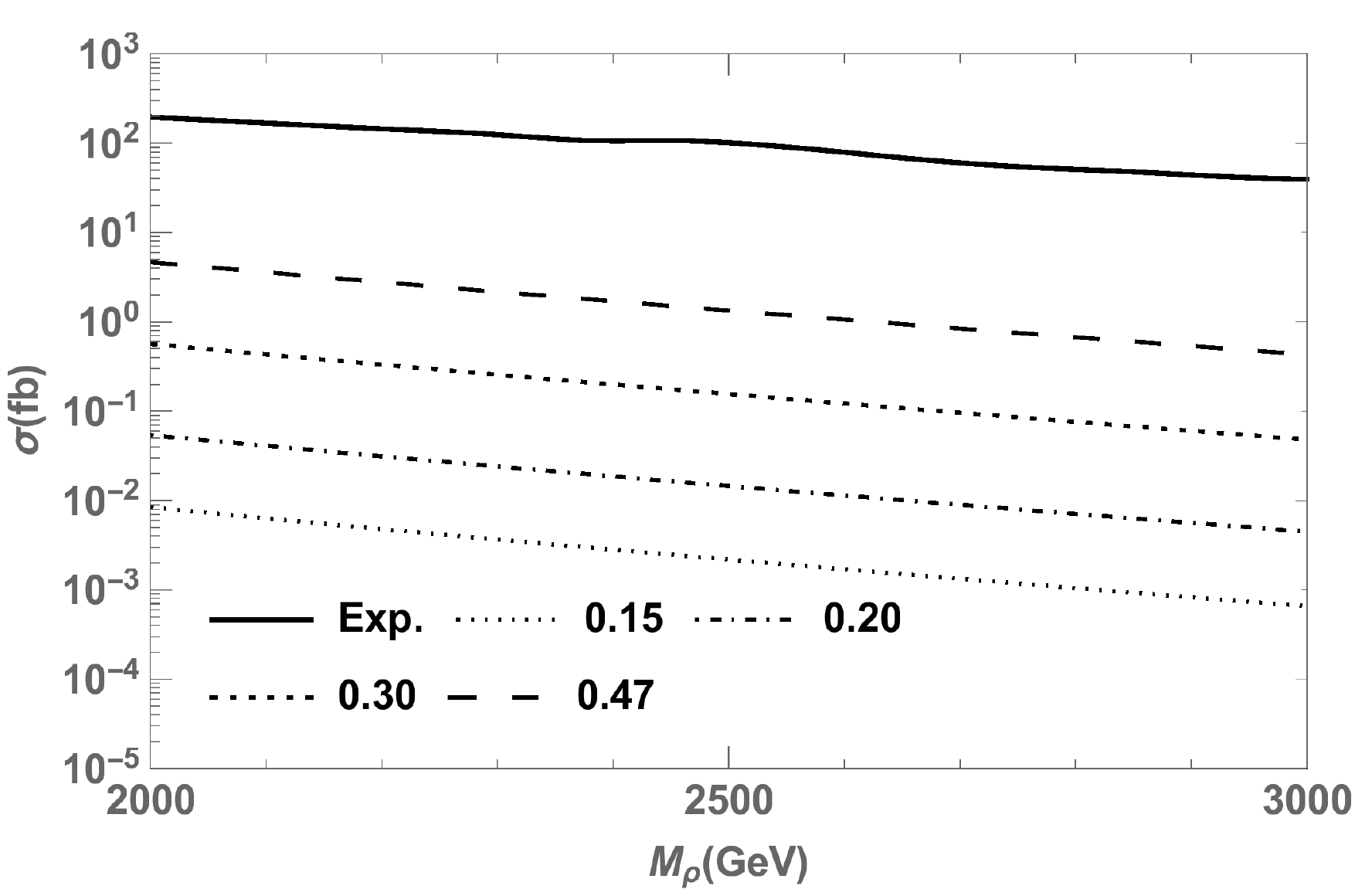}} 
\caption{Left: predicted  $pp\to \rho^{0,\pm} \to jj$ resonant production at $\sqrt{s}$ = 13 TeV as a function of $M_\rho$, for different values of $\tan\vartheta$; 
the solid line is the 95\% C.L. upper limit obtained  by ATLAS \cite{ATLAS:2016lvi}. 
Right: \emph{idem}, but for  $pp\to \rho^0\to t\bar{t}$; the solid line is the observed 95\% C.L. upper limit obtained by ATLAS \cite{TheATLAScollaboration:2016wfb}. 
}
\label{ULjj}
\end{figure}

We now focus on the LHC upper limits to constrain the model parameter space using the final states $l^+l^-$, $l\nu_l$ ($l=e,\mu$), $\tau^+\tau^-$, $jj$, $t\bar{t}$, $WZ$, $WW$, $WH$ and $ZH$, assumed to be produced through a resonant $\rho^0$ or $\rho^\pm$ decay. 
For example, the observation of the combined dilepton modes $e^+e^-$ and $\mu^+\mu^-$ \cite{ATLAS:2017wce}  provides a bound for a neutral resonance, which we identify here with the neutral state $\rho^0$. Fig. \ref{ULlep} (left) shows the cross section prediction for $\rho^0$ decaying into dileptons ($l=e,\mu$), together with the upper limit obtained by ATLAS, thus setting restrictions on the $\tan\vartheta$ and $M_\rho$ parameter space.
The CMS upper bounds in the $\tau^+\tau^-$ final state \cite{Khachatryan:2016qkc} are less restrictive than those of  $e^+e^-$ and $\mu^+\mu^-$, thus providing no further constraints as shown in Fig. \ref{ULlep} (right). The experimental bound in the $l\nu$ final state, with $l=e,\mu$, is as stringent as in the $l^+l^-$ channel (see Fig. \ref{ULlv}).  
In contrast to dileptons, the $t\bar{t}$ \cite{TheATLAScollaboration:2016wfb} and dijet \cite{ATLAS:2016lvi} experimental upper bounds do not restrict our parameter space, as can be seen in Fig.~\ref{ULjj}.  

\begin{figure}[h!]
\resizebox{17cm}{6cm}
{\includegraphics[scale=2.1]{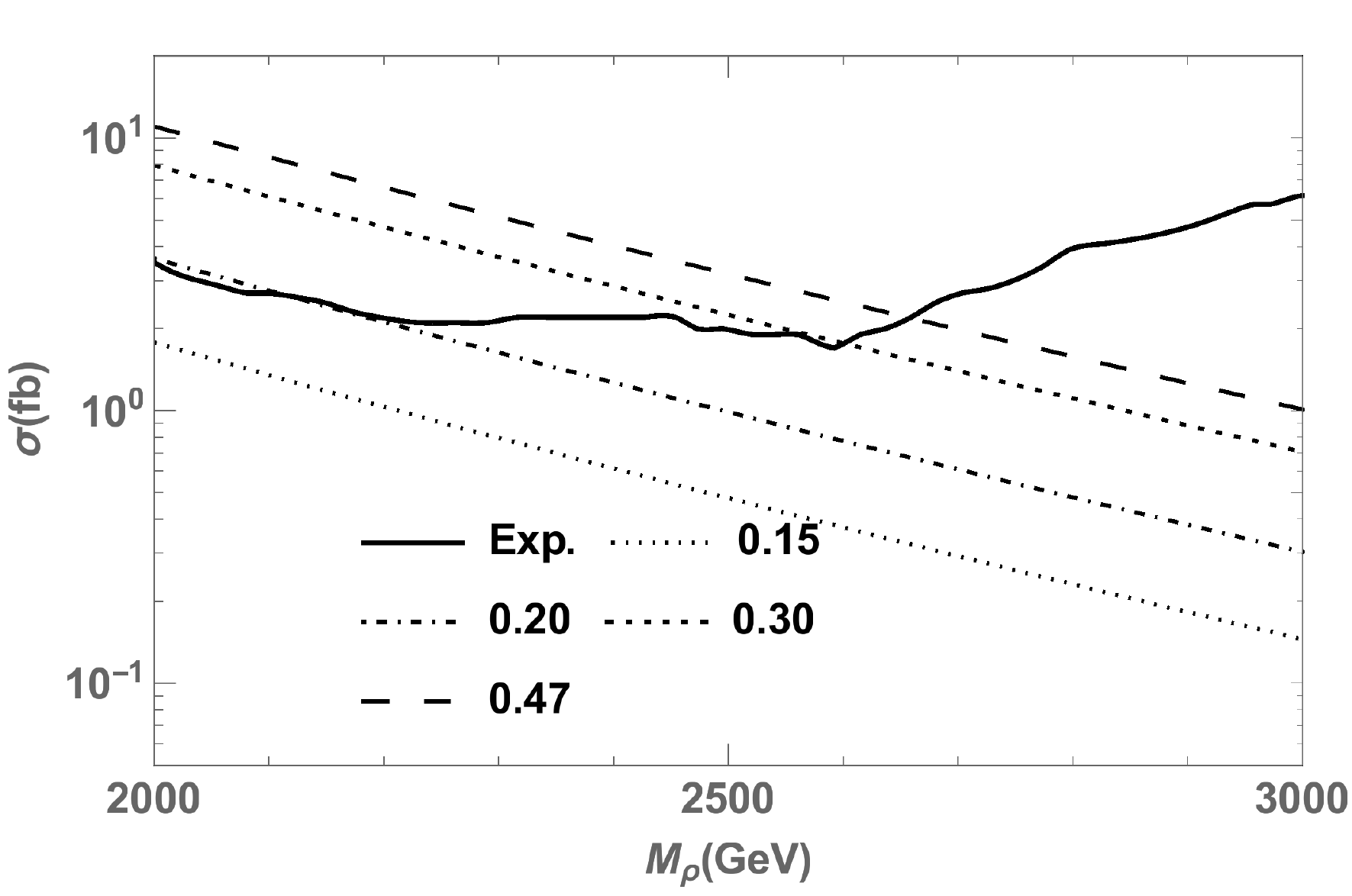}
\includegraphics[scale=2.0]{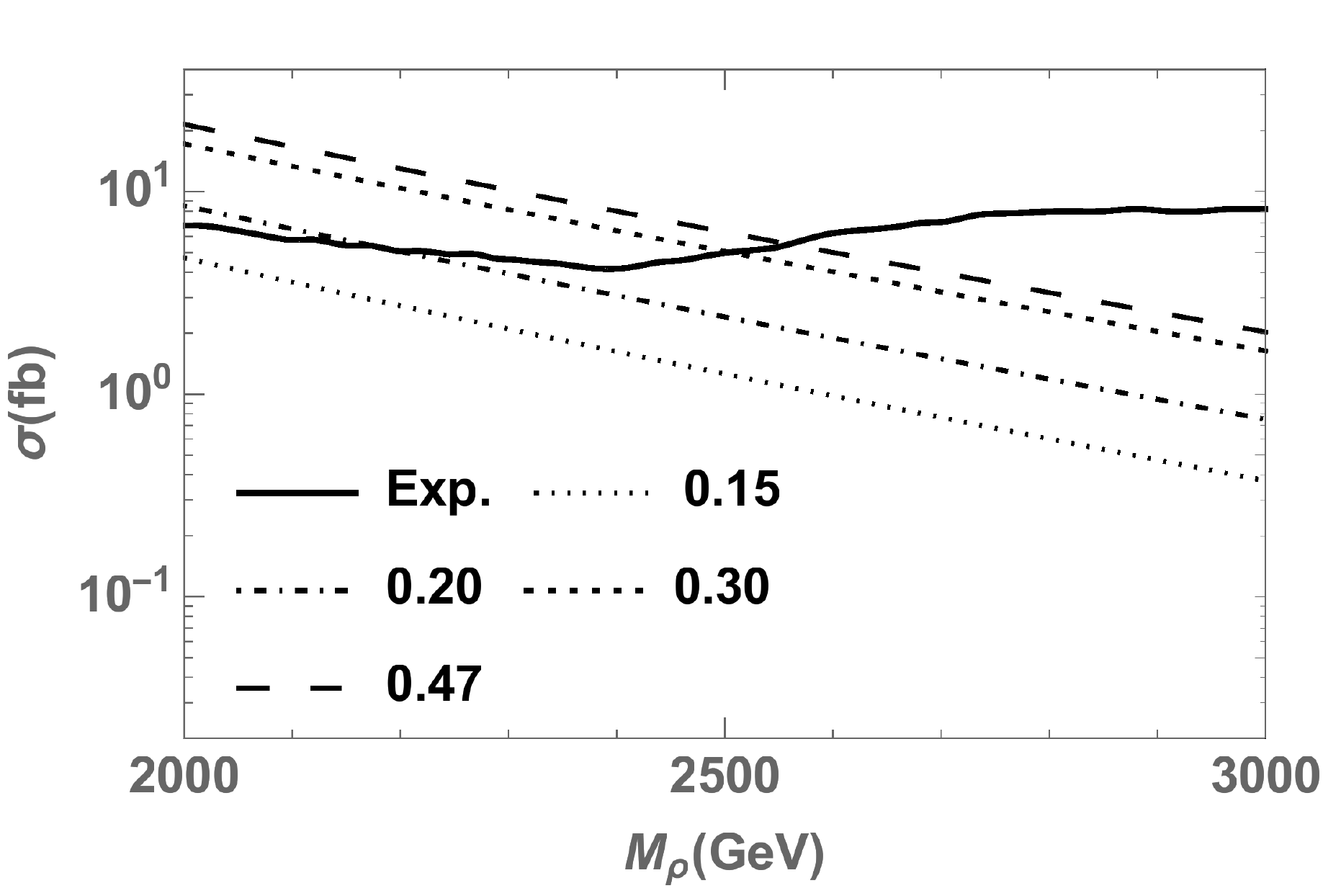}} 
\caption{
Left:  predicted $p p \to \rho^0\to Z H$ resonant production at $\sqrt{s}$ = 13 TeV  as a function of $M_\rho$, for different values of $\tan\vartheta$; the solid line is the observed 95\% C.L. upper limit obtained by ATLAS \cite{ATLAS:2017ywd}. 
Right: \emph{idem}, but for $pp \to \rho^\pm \to W^\pm H $.
} 
\label{ULZH}
\end{figure}

\begin{figure}[h!]
\resizebox{17cm}{6cm}
{\includegraphics[scale=1.95]{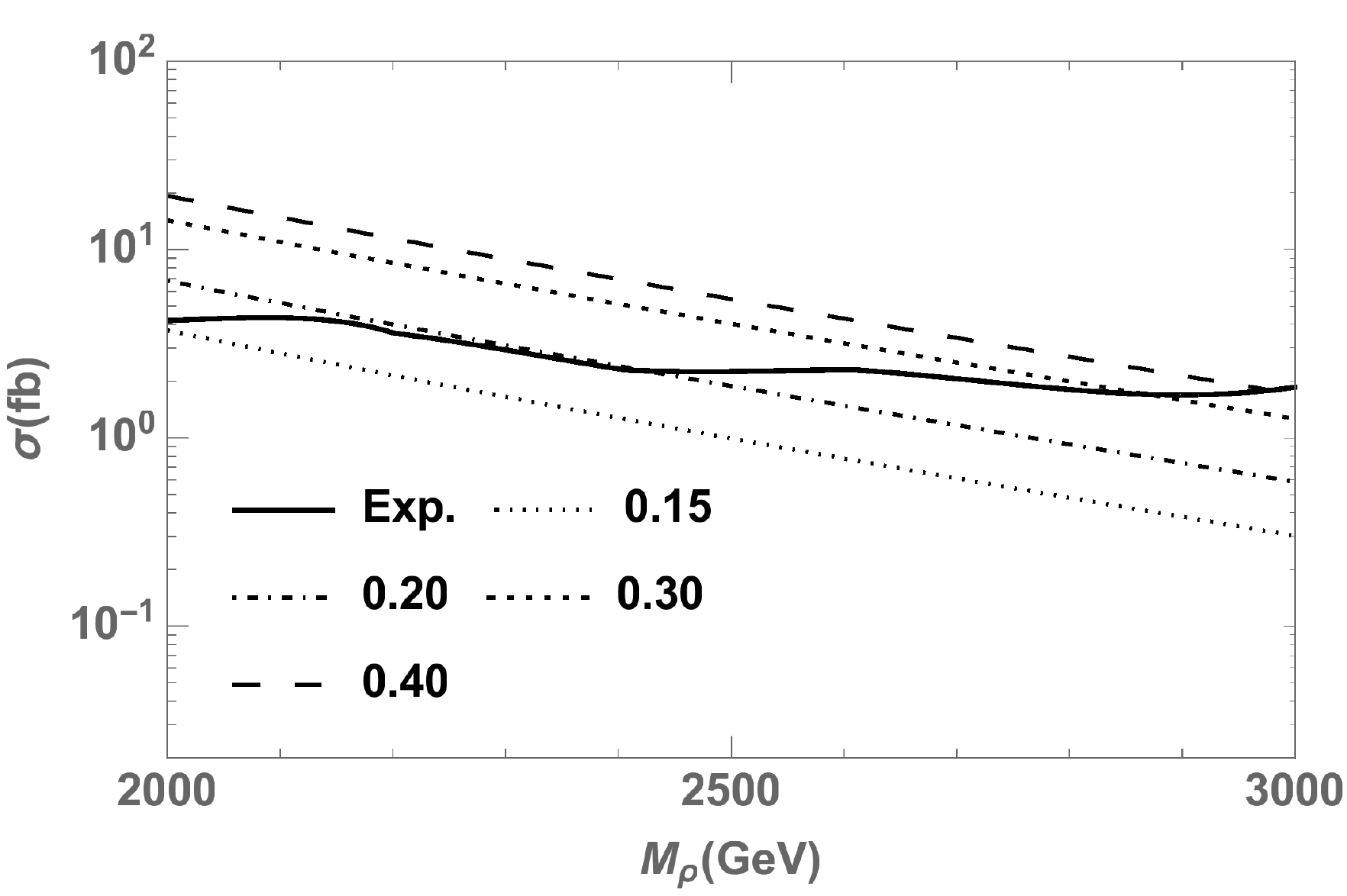}
\includegraphics[scale=1.90]{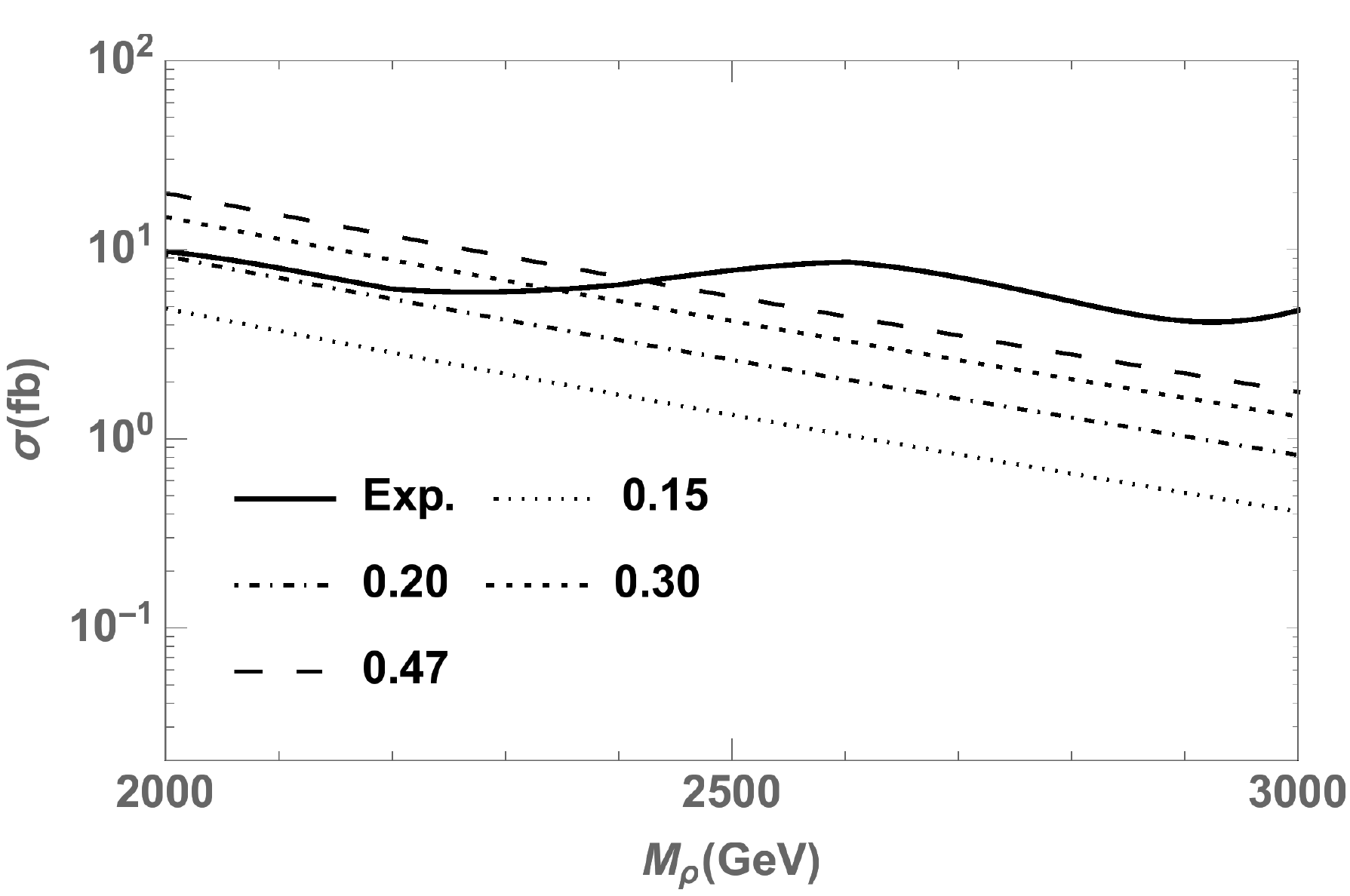}} 
\caption{
Left:  predicted $p p \to \rho^0\to W^+W^-$ resonant production at $\sqrt{s}$ = 13 TeV  as a function of $M_\rho$, for different values of $\tan\vartheta$;  the solid line is the  95\% C.L. upper limit obtained  
by ATLAS \cite{ATLAS:2016cwq}. 
Right: \emph{idem}, but for  $pp \to \rho^\pm \to W Z$.
}
\label{ULWW}
\end{figure}

\begin{figure}[h]
\resizebox{15cm}{8cm}
{\includegraphics{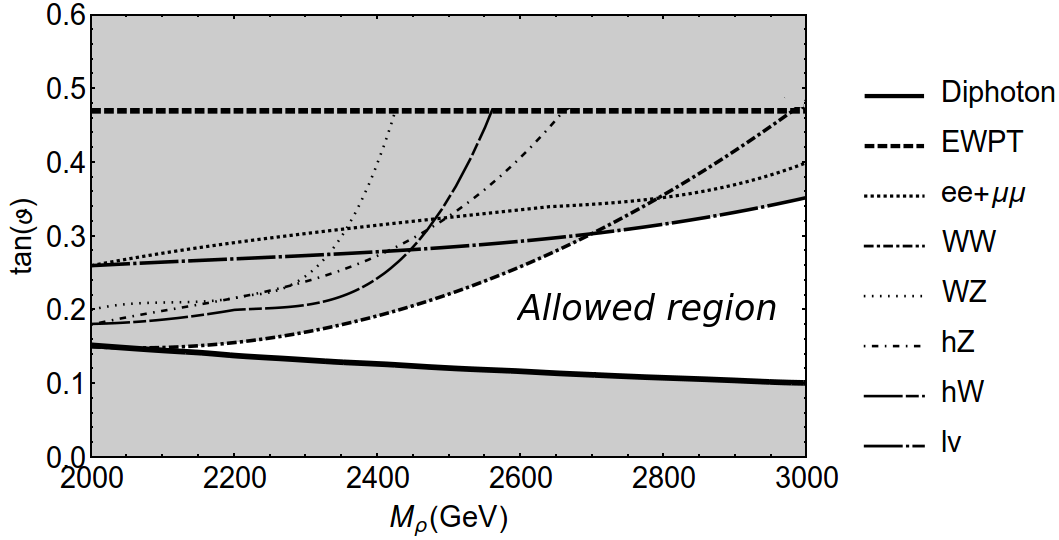}} 
\caption{Allowed and excluded regions in the model parameter space after including all the constraints, i.e. the $h\to \gamma\gamma$ (diphoton) constraint, the electroweak precision tests (EWPT) and the bounds from the LHC searches in the channels  $l^+l^-$,  $l\nu_l$ ($l=e,\mu$), $\tau^+\tau^-$, $jj$, $t\bar{t}$, $ZH$, $WH$, $WW$ and $WZ$. The allowed region which is consistent with all the constraints is shown in white, while the grey regions are excluded.}
\label{FPS}
\end{figure}

Now, the associated $h Z$ and $h W^\pm$ production estimates and upper bounds \cite{ATLAS:2017ywd} are shown in Fig.~\ref{ULZH}. 
Finally, the ATLAS constraints from $ZW$ and $W^+W^-$ production are shown in Fig.~\ref{ULWW}, where again we contrast the experimental upper bound \cite{ATLAS:2016cwq} with the resonant production of  $W^+W^-$ (left) and $ZW$ (right). 

Combining all these restrictions in the $\tan\vartheta$-$M_\rho$ plane, we arrive at Fig.~\ref{FPS}, where the allowed region of the parameter space is shown. Here we include also the $h\to \gamma \gamma$ (diphoton) constraint --which provides the lower bounds on $\tan\vartheta$, and the upper bound from electroweak precision tests (EWPT) --which turns out to be less restrictive than the upper bounds from the dilepton and diboson channels, as shown in the figure.

\section{Conclusions}

\label{conclusions}

We studied a framework of strongly interacting dynamics where the Higgs (a scalar doublet), and also a heavy vector triplet, 
appear as composite fields below a scale $\Lambda \simeq 4\pi v \sim 3$~TeV. 
Without addressing details of the strong dynamics, we focus on the  effective theory below the scale $\Lambda$,
assumed to be the Standard Model, with its $SU(2) _{L}\times U\left(1\right) _{Y}$ gauge group, with the addition 
of a  $SU(2) _{L}$ triplet of heavy vectors. 
The inclusion of the composite fields in the effective Lagrangian, 
 i.e. the Higgs and the heavy vectors,  is
done by considering the vectors as gauge fields of a hidden local $SU(2)_2$ symmetry, and the Higgs as a doublet under this same symmetry. On the other hand, the SM gauge group at this stage is a $SU(2)_1 \times U(1)$. The SM fermions transform only under the latter group. By the mechanism of hidden local symmetry, the  
$SU(2)_1\times SU(2)_2$ breaks down to the diagonal $SU(2)$ subgroup, which will be effectively the $SU(2)_L$ of the SM. This spontaneous breakdown is formulated in terms of a non-linear sigma model, where the ``would-be Goldstone'' bosons are absorbed into the massive vector triplets. 
In this process, the Higgs $SU(2)_2$ doublet, which originally interacts with the composite vectors only, now acquires interactions with the SM fields. In this way, the composite Higgs maintains a rather strong interaction with the composite vector triplets, and a weaker interaction with the SM fields.

 We put to test the resulting spectrum and interactions, in view of the existing experimental data: we determined the constraints arising from the measured Higgs diphoton decay rate, electroweak precision tests and the searches of heavy resonances at the LHC in the final states $l^+l^-$ and $l\nu_l$ ($l=e,\mu$), $\tau^+\tau^-$, $jj$, $t\bar{t}$, $WZ$, $WW$, $WH$ and $ZH$. 

As a consequence of these constraints, we find that heavy vector masses in the range $2.1$ - $3.0$ TeV are consistent with the data, together with a mixing of the heavy vectors with the SM gauge bosons in the range 
$\tan\vartheta\sim 0.1 - 0.3$.  These values are also consistent with the assumption that the Higgs couples weakly  to the Standard sector  and strongly to the heavy vector resonances. 
In other words, the current experimental data still allows for a Higgs boson that is strongly coupled to a composite sector, here assumed as triplet of vector resonances.

\subsection*{Acknowledgements}

We would like to thank F. Rojas, M. Schmaltz and J. Urbina for useful discussions. This
work was supported in part by Conicyt (Chile) grants ACT-146 and
PIA/Basal FB0821, and by  Fondecyt (Chile)
grants No.~1130617, 1170171, 1120346, 1160423 and 1170803. B.D was partially
supported by Conicyt Becas-Chile and PIIC/DGIP.

\end{document}